# CFD-CAA simulation of flow acoustic coupling in a half-dump combustor

**Ashoke De[*], AbhijitKushari, Sudharsan K, GeethaSriKonreddy**

Department of Aerospace Engineering, Indian Institute of Technology, Kanpur, India-208016

**Abstract:** This paper reports on the investigation of combustion instabilities in a Methane-air non-premixed half dump combustor for different flow Reynolds number (Re) using Computational aero-acoustic (CAA) simulation. In order to simulate the flow physics under turbulent flow conditions, Detached Eddy Simulation (DES) in conjunction with generalized eddy-dissipation model (EDM) is adopted, while a CAA formulation based on Lighthill's acoustic analogy is used for computing the acoustic field. It is observed that the unsteady pressure signals either predominantly arise from the natural acoustic modes of the duct or the local flow fluctuations in the vortex shedding process downstream of the dump plane, giving rise to different dominant frequencies at different spatial locations at lower Re and a single dominant locked-on frequency at higher Re. The non-dimensional numbers, i.e. Helmholtz and Strouhal numbers are used to characterize the duct acoustic modes from the vortex shedding modes. Under reacting flow conditions, unsteady heat release and pressure oscillations are monitored to compute Rayleigh index in order to verify if the instability is driving (positive net value) or damping (negative net value) at different frequency levels. Moreover, the predicted Helmholtz and Strouhal numbers are found to be in excellent agreement with the experimental data available in open literature for wide range of Re.

[*]Corresponding Author: Tel.: +91-512-2597863 Fax: +91-512-2597561

E-mail address: ashoke@iitk.ac.in

**Nomenclature**

| Symbols | Description |
|---|---|
| $C_p$ | specific heat capacity at constant pressure |
| c | speed of sound |
| D | molecular diffusivity |
| E | Total energy |
| e | internal energy |
| f | frequency |
| h | enthalpy |
| k | turbulent kinetic energy |
| $L_t$ | turbulent length scale |
| $l_x, l_y, l_z$ | length of the domain in 3 directions |
| $M_{w,i}$ | Molecular weight of species i |
| $n_x, n_y, n_z$ | order of modes in 3 directions |
| P | pressure |
| $Pr_t$ | turbulent Prandtl number |
| p' | pressure oscillation |
| q' | heat release oscillation |
| $Sc_t$ | turbulent Schmidt number |
| T | time period |
| $T_{ij}$ | Lighthill's tensor |
| $\bar{T}$ | temperature |
| t | time |
| $u_i$ | velocity vector |
| $x^+, y^+, z^+$ | non-dimensional wall distance in 3 directions |
| $\bar{Y}$ | Mass fraction |

Greek symbols

| | |
|---|---|
| $\Delta_s$ | local surface grid spacing |
| $\delta$ | boundary layer thickness |
| $\delta_{ij}$ | Kronecker's delta |
| $\mu$ | viscosity |
| $\mu_t$ | turbulent eddy viscosity |
| $\nu''_{j,r}$ | Stoichiometric coefficient for product j in reaction r |
| $\nu'_{i,r}$ | Stoichiometric coefficient for reactant i in reaction r |
| $\rho$ | Density |
| $\dot{\omega}_t$ | reaction rate |
| $\omega$ | Specific dissipation |

## 1. Introduction

Combustion Instability has been a subject of investigation for a long time [1-4], owing to the problems of noise generation and excitation of acoustic instabilities that leads to structural damage caused by enhanced vibrations or heat transfer in gas turbine combustors, boiler, furnaces, etc. One of the important driving mechanisms of combustion instability is primarily due to vortex shedding, as highlighted by Schadow and Gutmark [5], who reported that the development of coherent flow structures and their breakdown into fine-scale turbulence can lead to periodic heat release. Many workers have investigated dump combustors involving axisymmetric backward-facing step [5–8], that plays a predominant role in vortex shedding inside the combustor. In dump combustors, vortices are formed in the shear layer at the dump plane where the high and low speed streams interact with each other. As the flow experiences an adverse pressure gradient, the boundary layer is separated from the solid surface and after that it subsequently reattaches downstream to form a recirculation bubble. The recirculation bubble brings back the hot combustion product to mix with the fresh unburnt mixture and thereby forming the flame holding recirculation zone behind the dump plane. The vortex structure has a significant influence on the combustion process and has been successfully examined in different experimental facilities [9-11]. Different computational methods exist for predicting the combustion instability phenomenon in model/ practical combustors [12]. The methods range from simple reduced order models to solving full compressible Navier-Stokes equations. Most of the theoretical approaches decompose the components causing combustion instability, i.e. flow, flame and acoustics, and attempt to define/obtain the interaction between these components and predict the evolution of the overall system. Many investigators have used large-eddy simulation (LES) model directly to study combustion instability in practical combustors [13-21].

In the present investigation a CAA methodology (known as the hybrid approach), based on the Lighthill's analogy, is used for computing an acoustic field. The hybrid approach adopted in this work simulates the acoustic far field using a two-step procedure. In the first step, a turbulent flow field is computed in a flow simulation using a CFD method, which is capable of resolving the turbulent scales responsible for the noise generation. With the acoustic source information obtained from the first step of simulation, the far field sound pressure is calculated using the CAA formulation. Majority of the current simulation methodologies rely on this strategy, because a direct computation of both the sources and the resulting acoustic waves would be unaffordable for realistic practical problems due to plenty of reasons, as widely detailed in the literature [12, 22].

In order to simulate the flow physics under turbulent flow conditions, Reynolds-averaged Navier–Stokes (RANS) models are the most widely-used turbulence treatment approach in industrial applications. Nath [32] used unsteady Reynolds-averaged Navier-Stokes (URANS) approach in conjunction with a separate acoustics solver to capture the flow physics in the half-dump combustor and the predictions were unable to differentiate the hydrodynamic instability from the acoustic instability while capturing all the mode as acoustic instability for all the Re. Although offering greater accuracy, LES is always associated with greater computational cost, rendering the approach unfeasible in many situations. Recent research efforts have therefore focused on a family of hybrid

RANS-LES methods, intended to bridge the gap between these methodologies [23]. Of these, perhaps the most widespread is detached-eddy simulation (DES) [24], in which attached boundary layers are modeled using RANS and regions of separated flow resolved by LES. DES has been found to be a promising alternative to produce excellent results compared to RANS at a fraction of the cost of LES, particularly for flows dominated by large-scale separation.

Therefore, the primary objective of this present study is to capture the combustion-induced instability and unsteady vortex dynamics in a half dump combustor using hybrid turbulence model (DES) in conjunction with CAA methodology, while the primary emphasis is on the hydrodynamic instability at lower Re and acoustic instability at higher Re in order to differentiate the hydrodynamic mode of instability from the acoustic mode of instability which is often mixed-up in many literatures and commonly termed as acoustic instability as reported in [32]. Also, the focus is to establish the theory that hydrodynamic instability is primarily the driving mechanism at low Re due to unsteady vortex dynamics, which triggers the acoustic instability when moving to high Re regime due to strong coupling between heat-release and pressure fluctuations. Secondly, another goal of this work is to assess the predictive capability of reduced model for industrial applications. To this end a laboratory scale dump combustor for which extensive experimental database [9] exists, is considered for investigation. Furthermore, the unsteady heat release and pressure oscillations are monitored and local Rayleigh index is plotted to identify the driving mechanism of the instabilities. Also, Helmholtz and Strouhal numbers are analyzed to distinguish the natural acoustic modes and the vortex shedding modes for the varying Reynolds number in the combustor.

## 2. CFD-CAA: Numerical details

In the hybrid approach, the turbulent flow field is firstly computed using DES turbulence model, while CAA formulation is used in the next step to calculate the far field sound pressure based on the acoustic source information available from CFD simulations. The governing equations used for this CFD-CAA simulation are outlined below.

### 2.1 Governing equations for flow modelling

The Favre-filtered governing equations for the conservation of mass, momentum, energy and species transport are given as:

Continuity equation:

$$\frac{\partial}{\partial t}(\bar{\rho}) + \frac{\partial}{\partial x_i}(\bar{\rho}\widetilde{u_i}) = 0 \qquad (1)$$

Momentum equation:

$$\frac{\partial}{\partial t}(\bar{\rho}\widetilde{u_i}) + \frac{\partial}{\partial x_j}(\bar{\rho}\widetilde{u_i}\widetilde{u_j}) = \frac{\partial}{\partial x_i}(\bar{P}) + \frac{\partial}{\partial x_j}\left((\mu + \mu_t)\frac{\partial \widetilde{u_i}}{\partial x_j}\right) \qquad (2)$$

Energy equation:

$$\frac{\partial}{\partial t}(\bar{\rho}\bar{E}) + \frac{\partial}{\partial x_i}(\bar{\rho}\widetilde{u_i}\bar{E}) = -\frac{\partial}{\partial x_j}\left(\overline{u_j}\left(-\bar{P} + \mu\frac{\partial \widetilde{u_i}}{\partial x_j}\right)\right) + \frac{\partial}{\partial x_i}\left(\left(k + \frac{\mu_t C_p}{Pr_t}\right)\frac{\partial \bar{T}}{\partial x_i}\right) + \frac{\partial}{\partial x_i}\left(\bar{\rho}\sum_{s=1}^{N} h_s \left(D + \frac{\mu_t}{Sc_t}\right)\frac{\partial \bar{Y_s}}{\partial x_i}\right) \qquad (3)$$

Species transport equation:

$$\frac{\partial}{\partial t}(\bar{\rho}\bar{Y_i}) + \frac{\partial}{\partial x_j}(\overline{u_j}\bar{\rho}\bar{Y_i}) = \frac{\partial}{\partial x_j}\left(\left(D + \frac{\mu_t}{Sc_t}\right)\frac{\partial \bar{Y_i}}{\partial x_j}\right) + \overline{\dot{\omega}_i} \qquad (4)$$

In DES (hybrid RANS/LES) approach, the eddy-viscosity calculation depends on the grid-spacing in the computational domain. The inherent advantage of this approach is that it can be automatically switched to a sub-grid scale (SGS) model in the LES regions and to a RANS model in the RANS regions through the modification provided below.

The dissipation term of the turbulent kinetic energy is modified for the DES turbulence model as

$$Y_k = \rho \beta * k\omega F_{DES} \qquad (5)$$

where $F_{DES}$ is expressed as

$$F_{DES} = max\left(\frac{L_t}{C_{des}\Delta_{max}}, 1\right) \qquad (6)$$

where $C_{des}$ is a calibration constant used in the DES model and has a value of 0.61, $\Delta_{max}$ is the maximum local grid spacing ($\Delta_x, \Delta_y, \Delta_z$).

The turbulent length scale is the parameter that defines this RANS model:

$$L_t = \frac{\sqrt{k}}{\beta^*\omega}, \text{ where, } \beta = 0.09. \qquad (7)$$

The main practical problem with the DES formulation is that there is no mechanism for preventing the limiter from becoming active in the attached portion of the boundary layer. This will happen in regions where the local surface grid spacing is less than the boundary layer thickness, i.e., $\Delta_s < d\delta$, with $d$ of the order one [25]. In this case the flow can separate as a result of the grid spacing (grid-induced separation), which is undesirable. In order to reduce this risk, the DES-SST model also offers the option to "protect" the boundary layer from the limiter (delayed option). This is achieved with the help of the zonal formulation of the DES formulation, based on the blending functions of the SST model. The blending functions are functions of the wall distance.

$F_{DES}$ is modified according to

$$F_{DES} = max\left(\frac{L_t}{C_{des}\Delta_{max}}(1 - F_{SST}), 1\right) \qquad (8)$$

With $F_{SST} = 0, F_1, F_2$ where $F_1, F_2$ are the blending functions of the SST model.

In case $F_{SST}$ is set to 0, the original DES model is recovered. It should be noted that $F_{sst} = F_2$ offers the highest level of protection against grid-induced separation, but might also be less responsive in LES regions [25].

**2.2 Turbulence-chemistry interaction model**

The present work uses the eddy dissipation model (EDM) for turbulence-chemistry interaction closure where the reaction source term for each species in Eq. (4) gets modified. Eddy dissipation model is more suitable for reduced chemistry. The smaller among the expression (9) and (10) gives the limiting value for net rate of production $\overline{\dot{\omega}_{i,r}}$ of species $i$ due to reaction $r$,

$$\overline{\dot{\omega}_{i,r}} = v'_{i,r} M_{w,i} A \rho \frac{\varepsilon}{k} min_{\mathcal{R}} \left( \frac{Y_{\mathcal{R}}}{v'_{\mathcal{R},r} M_{w,r}} \right) \tag{9}$$

$$\overline{\dot{\omega}_{i,r}} = v'_{i,r} M_{w,i} AB \rho \frac{\varepsilon}{k} \frac{\sum_P Y_P}{\sum_j^N v''_{j,r} M_{w,j}} \tag{10}$$

Where,

$Y_P$ = the mass fraction of any product species, P

$Y_{\mathcal{R}}$ = the mass fraction of a particular reactant, $\mathcal{R}$

A = an empirical constant equal to 4.0

B = an empirical constant equal to 0.5

N = number of reacting species involved in the system

According to the eddy-dissipation model, the source terms are calculated based on the large eddy mixing time scale given by $k/\varepsilon$, where $k$ is the turbulent kinetic energy and $\varepsilon$ is the dissipation rate of turbulent kinetic energy [25]. The reaction (and hence combustion) occurs wherever $k/\varepsilon > 0$, which is the case in mixing controlled systems such as non-premixed flames. In the finite-rate/eddy-dissipation model, the source terms are calculated, using both the Arrhenius rate as well as the eddy-dissipation reaction rates. The smaller of the two values is used in the calculation of source terms. Thus, in mixing controlled region, the eddy dissipation model is turned on and in kinetically controlled regime, the laminar finite rate model takes over [25].

In the flow domain where LES equations are solved, the turbulent mixing rate, $\varepsilon/k$ in Eq. (9) and Eq.(10) is replaced by sub-gridscale mixing rate. It is given by

$$\tau_{sgs}^{-1} = \sqrt{2 S_{ij} S_{ij}} \tag{11}$$

Where,

$\tau_{sgs}^{-1}$ = subgrid-scale mixing rate (s$^{-1}$)

$S_{ij} = \frac{1}{2}\left(\frac{\partial u_i}{\partial x_j} + \frac{\partial u_j}{\partial x_i}\right)$ = strain rate tensor (s$^{-1}$)

In the present work the combustion is represented using methane-air global two step mechanism which includes six species ($CH_4$, $O_2$, $H_2O$, $CO_2$, CO and $N_2$) as tabulated below:

| Reaction | Pre-exponential factor | Activation energy (J/kgmol) |
|---|---|---|
| $CH_4 + 1.5O_2 \rightarrow CO + 2H_2O$ | $5.012 \times 10^{11}$ | $2.0 \times 10^8$ |
| $CO + 0.5O_2 \rightarrow CO_2$ | $2.239 \times 10^{12}$ | $1.7 \times 10^8$ |
| $CO_2 \rightarrow CO + 0.5O_2$ | $5.0 \times 10^8$ | $1.7 \times 10^8$ |

The first equation is irreversible, while the second and third reactions are reversible that lead to an equilibrium between CO and $CO_2$ in the burnt gases [21]. More details regarding EDM can be found in the literature [25-27].

## 2.3 Acoustics modelling using Lighthill's analogy (CFD-CAA)

Lighthill [28] was the first to propose this approach that offers viable alternatives to the direct method. In this method, the near-field flow obtained from flow modeling (URANS, DES, LES) is used to predict the sound with the help of integral solutions to the wave equations. Essentially, the acoustic analogy decouples the propagation of sound from its generation and allows the flow solution process to be separated from the acoustics analysis. In Lighthill's analogy, the momentum equation is rewritten while leaving out the viscous stresses as

$$\rho \left\{ \frac{\partial u_i}{\partial t} + u_j \frac{\partial u_i}{\partial x_j} \right\} = -\frac{\partial P}{\partial x_i} \qquad (12)$$

Equation(9) can be written in conservative form as

$$\frac{\partial \rho u_i}{\partial t} + \frac{\partial \rho u_i u_j}{\partial x_j} = -\frac{\partial P}{\partial x_i} \qquad (13)$$

Eliminating $\rho u_i$ and subtracting $c_0^2 \frac{\partial^2 \rho'}{\partial x_i \partial x_j}$ in Equation(10) gives the Lighthill's equation

$$\frac{\partial^2 \rho'}{\partial t^2} - c_0^2 \frac{\partial^2 \rho'}{\partial x_i \partial x_i} = \frac{\partial^2 T_{ij}}{\partial x_i \partial x_j} \qquad (14)$$

where $T_{ij} = \rho u_i u_j + (P - c_0^2 \rho)\delta_{ij}$

Equation (14) is analogous to wave equation with a source term. The RHS of Equation (14) is the source term, which is evaluated using DES formulation. The LHS of Equation (14) is the propagation term. More details about the Lighthill's analogy can be found in literature [25, 28].

## 3. Computational domain and boundary conditions

The geometry chosen for the simulations is a three dimensional half-dump combustor [8]. Figure 1(a) shows the three-dimensional schematic view of the flow domain used in the present simulations. The computational domain consists of a streamwise length $L_x$= 48$h$, including an inlet section $L_i$ = 14.67$h$ prior to the sudden expansion, vertical height $L_y$= 2$h$ and spanwise width $L_z$=2$h$, where $h$ is the step height (taken as 30 mm in the present work). The coordinate system is placed at the lower step corner as shown in Fig. 1(a). A hexahedral structured mesh is used for the flow domain. Three different mesh sizes are used for the flow simulations using the DES model. Uniform grid spacing is selected for the streamwise and spanwise directions. In the vertical direction, a non-uniform mesh distribution is used with fine grid spacing near the lower wall and at the step. The finer mesh consists of 936×91×14 grid points and the medium mesh consists of 624×61×9 grid points while the coarse mesh consists of 312×31×5 grid points. The simulations are performed using ANSYS FLUENT 13.0 [25]. The pressure and velocity coupling has been evaluated using the SIMPLE (Semi-Implicit Method for Pressure Linked Equations) algorithm and a second order upwind scheme has been used to compute the convective fluxes of all the equations except momentum. The momentum equation is evaluated using a second order bounded central difference scheme, whereas the time is discretized using a second order implicit scheme.

The mean inflow velocity profile, $U(y)$, imposed at the left boundary $x = -L_i$ is a turbulent boundary layer profile with $U_o$ being the maximum inlet velocity. The step-height Reynolds number is defined as $Re_h = U_0 h/\nu$, and the expansion ratio as $ER = L_y/(L_y - h)$. In the present simulations, the expansion ratio ($ER = 2$), boundary layer thickness at the step ($\delta_{99}/h = 1.2$) and Reynolds number ($Re_h=18000$) are considered as per measurements reported in the literature [9]. In the spanwise direction, the flow is assumed to be statistically homogeneous and periodic boundary conditions are applied at Z=0 and Z=$L_z$. No-slip boundary conditions are used at all other walls. The mean velocity profile, $U(y)$, obtained from 1/7th power law, is used at the inlet while the pressure outlet is used at the outlet plane. The time step in the current simulation is chosen using the CFL (Courant-Freidrich-Lewy condition) criteria. Once the initial transients are over (~9 flow-through times), the statistical analyses are performed for large number of samples (~15 flow through times) in order to achieve good statistical data.

For the reacting case, fuel is injected at the step plane through a square port of width equal to 0.13 times the step height, located at 1 mm below the step corner (Fig. 1(b)) and centered in the domain in the spanwise direction. A constant mass flow rate $\dot{m}_f$ for fuel of 142 mg/s is specified at the fuel port, similar to the experiments [9]. Detached eddy simulation (DES) is used for modelling the turbulent flow-field while Eddy Dissipation Model (EDM) [25-27] is used for modelling the turbulent chemistry interaction. It should be noted that the inlet boundary condition is specified at 14.67$h$ upstream of the step plane in order to ensure that the effect of error (e.g. prescription of random velocity superposed over the mean velocity to approximately account for the turbulence in the incoming flow) due to the inlet condition remains minimal at the step plane. It also allows for shear layer dynamics to affect the flow in the inlet part of the domain as observed in experiments [33] during active flame acoustic lock-on. Figure 1(c) shows a cross sectional view of the backward facing step (BFS) with two receiver points, which are located at $X_p/L$= 0.35 and 0.83 where $X_p$ is the distance measured from the inlet and L is the length of the combustion chamber from inlet, are considered for monitoring the unsteady pressure in the combustion chamber (Fig. 1(b)) to maintain consistency with the measurements in reference [9].

### 4. Results and Discussion

We will first present the cold flow simulations to examine the choice of mesh along with the turbulence model. Then, the reacting flow simulations are reported with the chosen mesh by specifying a fuel mass flow rate of 142mg/s. An isothermal case study is performed only for $Re_h$ = 18000, whereas coupled simulations are performed for a range of Reynolds numbers from 14000 to 47000.

**4.1 Grid independent study & Non-reacting flow results**

The grid independent study has been carried out for three different sets of grids as noted and mentioned earlier. Figure 2 depicts the profiles of mean axial velocity and rms of axial velocity at different stream wise locations for different meshes. The reattachment length is also compared and shown in Fig. 3. Due to the increase in shear layer thickness, the fluctuation spreads with increase in axial distance. The length of the recirculation zone is determined as the axial distance from the step where the $U = 0$ line in the measurement plane intersects with the bottom wall. As observed from the plots, all the meshes produce similar profiles while the very fine mesh and fine mesh predictions are in excellent agreement with each other and with the measurements as well. However, the coarse mesh predictions are not in good agreement with the measurements. Furthermore, predicted reattachment length by the coarse mesh is found to be x/h=6.16 (h=step height) while the predictions using other two meshes are found to be x/h=6.03which are in excellent agreement with the experiments. The fine mesh having the grid spacing in the three directions in wall units (near-wall layer scales are denoted by wall units, they are defined in terms of friction velocity $u_\tau$ and kinematic viscosity $\nu$ and it is denoted by a '+' superscript) as $x^+= 2$, $y^+_{min}= 0.14$, $y^+_{max}=3$, and $z^+ =6$, respectively, are found to be adequate for capturing the flow physics. Hence, the fine mesh has been chosen for the rest of the flow simulations.

**4.2 CAA results**

In this section, we report the results of CAA analysis for a non-premixed half-dump combustor, in which methane is injected at the backward-facing step, which mixes and burns with the air flowing past the step in the unsteady recirculation zone. The flow field simulation is done with DES turbulence model and fluctuating components (i.e. fluctuating pressure and heat release) is provided as source terms for CAA analysis. The flow Reynolds number is varied over a range to gradually change from conditions of low-amplitude noise to excitation of high amplitude discrete tones.

Figures 4(a) and 4(b) exhibit the dominent frequencies observed in the acoustic spectra measured at two different locations from the backward facing step for multiple Reynolds numbers in terms of Helmholtz number *(He)* and Strouhal number *(St)*. The Helmholtz number is defined as *He=fL/c*, where *f* is observed frequency, *L* is the length of the combustor, and *c* is the speed of the sound at the reference condition, i.e. at the inlet in this case. Whereas the Strouhal number is defined as *St=fh/U*, where *h* is the height of the backward facing step and *U* is the average airflow velocity at the inlet. The Helmholtz number remains relatively constant with respect to the flow Reynolds number if the observed frequency is that of the natural acoustic mode of the duct. On the other hand the Strouhal number remains constant relative to the flow Reynolds number if the observed frequency is that of the vortex shedding frequency in the recirculation zone downstream of the backward facing step. Helmholtz number increases linearly with the airflow Reynolds number if the observed frequency is the vortex shedding frequency and the Strouhal number decreases hyperbolically with Reynolds number if the observed frequency is that of natural acoustic mode of the duct [31]. The data in Figs. 4(a) and 4(b) show that both the *He* and *St* are constant at higher Re, whereas *St* approaches a constant value after decreasing hyperbolically in the lower Reynolds number range. This behavior of a nearly constant Helmholtz number at higher Re transitioning to a linearly increasing trend at lower Re (or a hyperbolic trend in Strouhal number asymptoting to a constant value) and the corresponding nonlinear rise in the amplitudes with increase in the flow Reynolds number is due to the flow-acoustic lock-on, where the dominent frequency observed is due the vortex shedding in the shear layer down stream of the backward facing step. Typically, the step mode vortex shedding frequency gives rise to the Strouhal number of 0.2, which is reported as the frequency at which the flow-acoustic lock-on [9]. As observed, the predictions nicely capture both the trend and are in excellent match with the measurements. The present predictions uses highly resolved mesh along with DES methodolgy that allow to capture different modes accurtely, therby making these predictions superior to earlier predictions reported in literature [32] where the authors used URANS and reported all the modes as acoustic modes at all the Re. While, in the present study, at lower Re, two distinct modes are observed but a single mode is exhibited at higher Re when the hydraudynamic instability couples with acoustics instability as reported in [31].

Figure 5 exhibits the phase average plots of velocity and temperature at the center of recirculation bubble for five different time instances representing different phases during a complete cycle of the most dominant frequency observed in the frequency spectra for Re= 18000, 26000, 36000 and 47000, and the data at each phase is averaged over four of these cycles in order to eliminate some of the turbulent fluctuations arising due to higher frequency oscillations. The cycle shows a repeatable pattern at phase5 ($360^0$) starting with phase1 ($0^0$) representing a shift in the vortex bubble center in one cycle of oscillation, which is one of the important driving mechanisms of combustion instability in the present case. These shear layer oscillations downstream of the backward-facing step modulate the heat-release fluctuations that are the source of acoustic oscillations, and the frequency of the former oscillations matches one of the natural acoustic modes of the duct. The rise and the subsequent fall of the amplitudes are symptomatic of resonance between these two processes. The high amplitude pressure oscillations are excited when the periodic heat release and pressure oscillations are in proper phase. This may lead to the excitation of acoustic oscillations, and that can be sustained by addition of heat as suggested by Rayleigh [34]. Thus, the excitation is maximum when the heat and pressure oscillations are in phase with each other. One of the most widely used form of the Rayleigh's criterion in literature is the integral form and is evaluated using the local Rayleigh Index (*RI*) over one cycle of instability as shown below [35]:

$$RI = \frac{1}{T}\int_{t}^{t+\frac{2\pi}{\omega}} p'q' dt \qquad (15)$$

where p' and q' are the pressure and heat release oscillation, T is the time for one cycle of instability and ω is frequency in rad/s. Usually, a positive *RI* indicates an amplification of the pressure oscillation due to the fluctuating heat release rate, whereas a negative *RI* denotes a dampening of the pressure oscillations.

In the present case, the modes of dominant acoustic frequencies observed in the domain are calculated as $f = \frac{c}{2}\left[\left(\frac{n_x}{l_x}\right)^2 + \left(\frac{n_y}{l_y}\right)^2 + \left(\frac{n_z}{l_z}\right)^2\right]^{1/2}$, where $n_x, n_y, n_z$ are the order of modes, $c$ is the speed of sound and $l_x, l_y, l_z$ are the length of the domain in 3 directions. Since, the transverse modes in y & z directions are not present (1st mode in 'y/z' is ~2900 Hz under cold condition, i.e. T=300K), the flow is dominated by longitudinal modes only and represented by $f = \frac{c}{2}\left(\frac{n_x}{l_x}\right)$. Initially, the predictions for $Re_h$ = 47000 is discussed in details by studying both the reacting and non-reacting cases to illustrate the mode of instability and driving mechanism, and then followed by the results of other Re. Figures 6-9 show results for flow Reynolds number 47000. Figures 6(a) and 7(a) exhibit the power spectral density with respect to frequency at two different locations. The sampling rate for power spectral density estimation is 50 KHz ($\Delta t=2 \times 10^{-5}$ s) and the total sample size and block size are 42494 and 512, respectively used for the given Re. It could be seen that multiple dominant frequencies at 149, 711, 860, 1006, 1572, 1721, 1867, 2581, 3442 and 4302 Hz exist in the domain (frequency range 5000Hz). The presence of multiple peaks in the pressure measurement shows that there is a feedback between pressure and heat release rate oscillations at the acoustic frequency for this Re, because the observed modes are either harmonics or superposition of two different modes. In order to determine if the periodic heat release is a driving force, it is necessary to have a complete knowledge of heat release and acoustic pressure in the combustion chamber. Figures 6(b) and 6(c), corresponding to pressure and heat release rate fluctuations at the vortex shedding region ($X_p/L$=0.35), are showing most correlated frequency of oscillations to be at 148.8 Hz, while Figs. 7(b) and 7(c) correspond to pressure and heat release rate fluctuations at the downstream location ($X_p/L$=0.83) and exhibit the similar behavior. This most correlated frequency (i.e. beat frequency) is arising due to the movement of the shear layer as mentioned earlier, i.e. motion of recirculation bubble center as shown in Fig. 5. The other peaks at 860, 1721, 2581, 3442, and 4302 Hz are the 2nd, 4th, 8th and 10th acoustic modes of the combustor without inlet (based on the combustion chamber length downstream of the step), while the 711, 1008, 1572, and 1869Hz are arising purely due to the superposition of the natural modes and lock-on acoustic mode (148.8 Hz), e.g. 711 Hz and 1008Hz are nothing but (860±148.8 Hz) and similarly 1572 Hz and 1869 Hz are (1721±148.8 Hz). Furthermore, the flow field for reacting flow is seen to closely resemble the cold flow field, suggesting strong influence of vortex dynamics on combustion process. Figure 8 shows frequency spectrum obtained for Re=47000 from cold flow simulation using CAA methodology. Multiple peaks of comparably very low amplitudes are observed at location 0.35 in the vortex shedding region whereas at location 0.83 a frequency of 148.8Hz is observed, which signifies the excitation of a natural acoustic mode without lock on to the vortex shedding in the shear layer. Regardless of the multiple peaks, even when they are of comparable amplitude, the most dominant frequency in the observed amplitude spectra is selected under each flow condition for the cold flow analysis. It is also observed that the vortex formed downstream of the step-face exhibits a cyclic behavior with the dominant frequency due to recirculation and flow reversal. The reacting flow field temperature and velocity phase averaging along the stream-wise direction also support this observation by showing similar behavior of vortex in the combustor (Fig. 5).

Moreover, it is also observed that heat release is in phase with the pressure oscillations and the net value of the local Rayleigh index, estimated for different frequency (cycles), suggests that driving (positive net value) occurs at 148.8 Hz as shown in Fig.9. The harmonics of other dominant frequencies also exist but are of lesser magnitude compared to 148.8 Hz. Figure 9 shows the Rayleigh index obtained using Equation (15) for the observed dominant frequencies in the domain marking the onset of instability at 148.8 Hz, signifying the excitation of natural acoustic mode with lock-on to the vortex shedding in the shear layer leading to instability. The CAA data shows that combustor flow for this Re is dominated by vortex flow in the flame holding region and associated with periodic heat release. Based on the CAA database, the origin of these instabilities are attributed to the key event such as vortex shedding which is modulating the heat release rate fluctuations in the recirculation zone downstream of the dump plane, followed by unsteady combustion in the oscillating shear layer of that zone.

Similar exercise is carried out for Re=36000 and the results are depicted in Figures 10-13. As expected, there are multiple dominant frequencies 95.9,685,1373,2059 Hz that exist in the domain for a frequency range up to 2500 Hz (Figs. 10(a) and 11(a)). The most significant peak at 95.9 Hz corresponds to the oscillation of the recirculation bubble as noted and mentioned earlier (Fig. 5) and the other frequencies (685, 1373, 2059) are the $2^{nd}$, $4^{th}$, and $6^{th}$ longitudinal acoustic modes of the combustor with inlet. The process is a resonant one and the pressure grow because of a resonant acoustic mode being fed at an appropriate phase (Rayleigh index) as seen from Figs. 10(c) and 11(c). Fig. 13 shows the Rayleigh index obtained for the dominant frequencies in the domain marking the onset of instability at 95.9 Hz. Figure 12 shows frequency spectrum for the cold flow where multiple peaks of comparably very low amplitudes are observed at location 0.35 in the vortex shedding region whereas a frequency of 99Hz is observed at location 0.83, which signifies the excitation of a natural acoustic mode without lock on to the vortex shedding in shear layer. Similar to Re=47000, the flow field for reacting flow is seen to closely resemble the cold flow field such that some of the dominent frequencys appeard in the reacting flow also exist in the non-reacting simulations.

The above-mentioned cases are the situations where pressure and heat-release are in phase and produce the lock-on acoustic mode. In order to analyze the behavior at lower Re, we repeat the similar exercise for Re=26000 and 18000. Figures 14-17 depict the predictions for Re=26000. From Figs. 14(b) and 15(b), it could be seen that the combustor flow, which is dominated by vortex dynamics in the flame holding region, is not associated with periodic heat release. As observed, the most dominant frequencies are found to be 101 Hz and 99 Hz at locations 0.35 and 0.83, which correspond to the modes due to hydrodynamic instabilities associated with vortex shedding (Figs. 4 and 5). The first one (99Hz) is primarily due to the motion of recirculation bubble center while the $2^{nd}$ one (101 Hz) arises due to superposition of two waves of different frequencies, i.e. 99Hz and the bulk-mode frequency of the inlet section (195 Hz). The process is not a resonant one and the pressure fluctuations do not diverge because the resonant acoustic mode is not being fed at an appropriate phase (Rayleigh index) as seen from Fig.14(c) and 15(c). Fig. 17 shows the Rayleigh index obtained for the dominant frequencies in the domain, which shows a lack of acoustics instability leading to a conclusion that the dominant oscillations observed are primarily hydrodynamic instabilities and thereby indicating no strong correlation with acoustics modes. Since the heat release oscillations are not in phase with the pressure oscillations the net value of the integral of the curve (pressure and heat release rate measurements at 0.35 and 0.83 with respect to time) exhibit that damping (negative net value) occurs for Re=26000. Furthermore, the flow field for reacting flow is seen to closely resemble the cold flow field as shown in Fig. 16. Some of the doment frequencies that appear in the reacting flow (e.g., 101 and 99 Hz) also exist in the non-reacting simulations suggesting that they are purely hydrodynamic in nature.

Moving to Re=18000, it can be observed from Figs. 18(a) and 19(a) that most dominant frequencies appear to be 46 Hz and 139.8Hz at locations 0.35 and 0.83, signifying the excitation of hydrodynamic mode without lock on due to the vortex shedding in the shear layer (Figs. 4 & 5). Similar to Re=2600 case, the first one (46Hz) is primarily due to the motion of recirculation bubble center while the $2^{nd}$ one (139.8 Hz) arises due to superposition of two waves of different frequencies, i.e. 46Hz and the bulk-mode frequency of the inlet section (195 Hz). As mentioned and noted earlier, for this case also, the process is not a resonant one and the pressure does not diverge as a resonant acoustic mode is not being fed at an appropriate phase (Rayleigh index) as seen in Figs.18(c) and 19(c). Figs. 18(b) and 19(b) confirm the observation that there is no feedback between pressure and heat release rate at the acoustic frequency for this Re. Moreover, Fig. 21 shows the Rayleigh index obtained for the dominant frequencies in the domain suggest acoustically stable combustion concluding that the dominant oscillations observed are primarily hydrodynamic instabilities. Furthermore, the presence of the doment frequenies, i.e. 46 and 139.8 Hz, in both reacting and non-reatcing cases, confirms that they are purely hydrodynamic in nature and not associated with acoustic instability which happens to appear at higher Re.

All the simulations show that the instability is associated with the vortices in the mixing layer, which couples with the acoustic pressure to excite strong oscillations. When the non-reacting flow is forced by the upstream or downstream duct resonant acoustic modes, vortices can be generated at much larger scale with the flow velocity, i.e. at the larger Re. However, since the acoustic emission of non-reacting vortices is low, there is no detectable feedback between flow and acoustic pressure in the chamber. Alternatively when a reacting mixing layer is considered, the large periodic energy release associated with the burning along the shear layer provides the missing link for the necessary feedback loop. The energy release is thus periodic in nature and reaches a maximum when the vortices breakdown into small-scale turbulence. Thus, the fluctuating heat release can feed back energy into the acoustic pressure at high Re leading to a lock-on situation as opposed to the low Re cases.

## 5. Conclusions

In this study, the hybrid CAA methodology is used to investigate a laboratory scale half dump combustor in order to gain additional insight into combustion instabilities and unsteady vortex dynamics. To simulate the turbulent flows, DES model is used and quantitative comparison is made with the experimental data in order to verify the accuracy of the DES model. Later on, DES model in conjunction with generalized eddy-dissipation model (EDM) is adopted for reacting calculations where unsteady pressure and heat release oscillations are monitored and combined to obtain information about the flow feature using CAA methodology. The database obtained using CAA methodology is used for studying the basic physical phenomena and Rayleigh index at different frequency levels. The simulations show that the instability is associated with the vortices in the mixing layer, which are coupled with the acoustic pressure to excite strong oscillations. The fluctuating heat release is found to feed energy in to the acoustic pressure oscillations at high Re leading to a lock-on situation as opposed to the low Re cases. Moreover, the predicted Helmholtz and Strouhal numbers are found to be in excellent agreement with the measurements for a wide range of Re. Hence, it can inferred from the present simulations that the hybrid CAA methodology has the potential to investigate fundamental physical phenomena, e.g. combustion instability, vortex dynamics etc., for large scale practical problems with greater accuracy at reasonable computational cost.

**Acknowledgments**

Simulations are carried out on the computers provided by the Indian Institute of Technology Kanpur (IITK) (www.iitk.ac.in/cc) and the manuscript preparation as well as data analysis has been carried out using the resources available at IITK. This support is also gratefully acknowledged.

(a)

(b)

(c)

Figure 1. (a-b): Schematic view of the computational domain, (c) Acoustic monitoring points at mid-span plane at near step location Xp/L=0.35 and downstream location Xp/L=0.83, where $X_p$ is the distance measured from the inlet

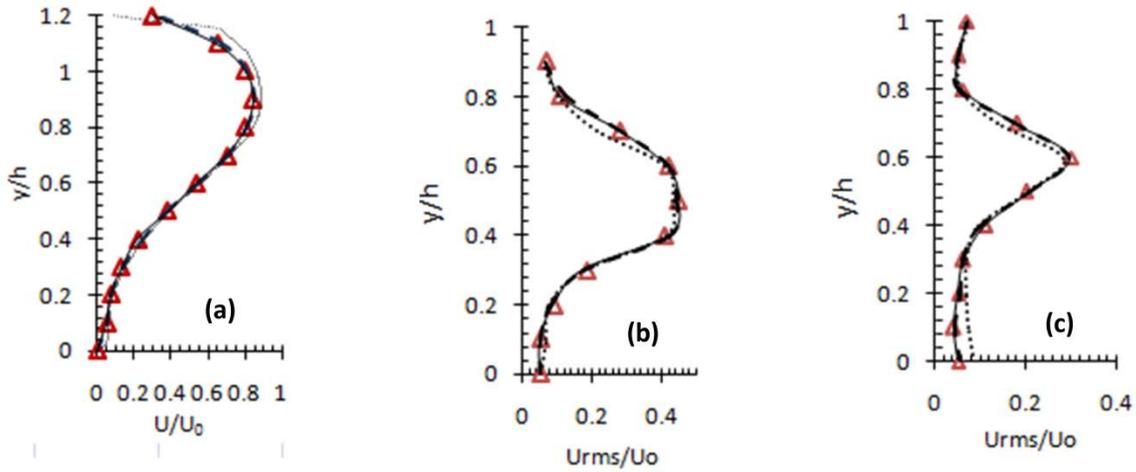

Figure 2. Non-reacting flow results for Re= 18000 at different streamwise locations h [Step height]: Experimental data(Δ)[9], lines are DES predictions: finest mesh ( ──),fine mesh(- - - ), coarse mesh (…..). (a) Mean stream wise velocity $U/U_0$ at x/h = 6, (b) streamwise velocity fluctuation $U_{rms}/U_0$ at x/h = 2,(c) streamwise velocity fluctuation $U_{rms}/U_0$ at x/h =7

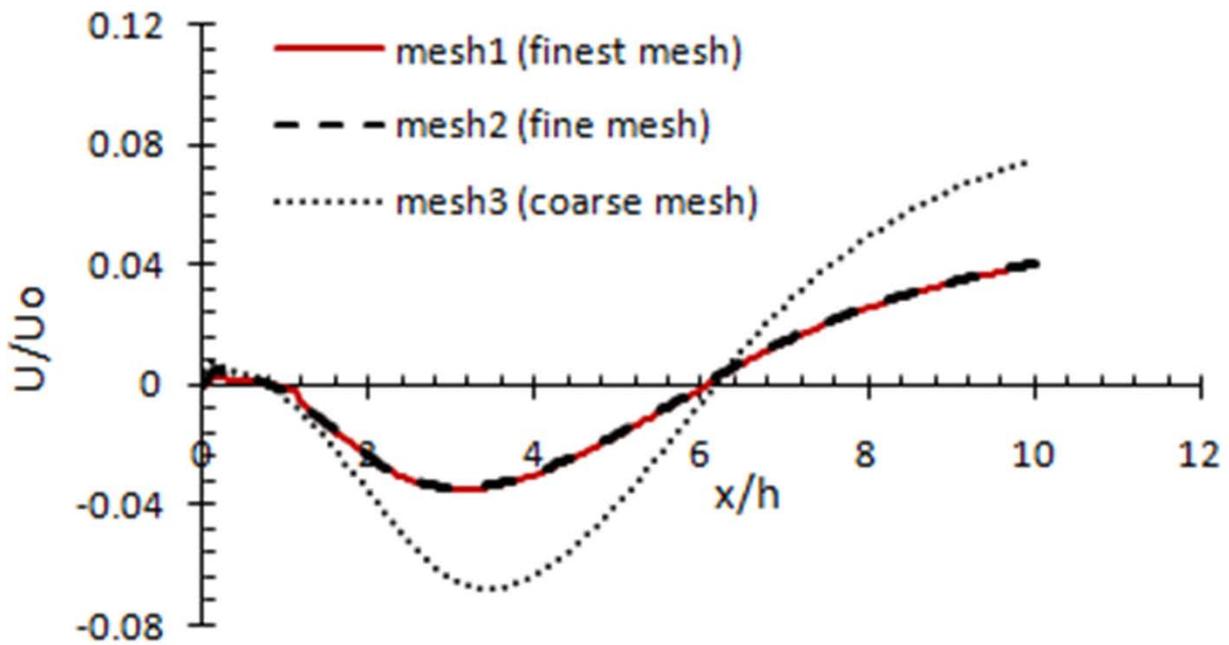

Figure 3. Recirculation Length Xr :Streamwise Velocity at first cell above wall for Re=18000

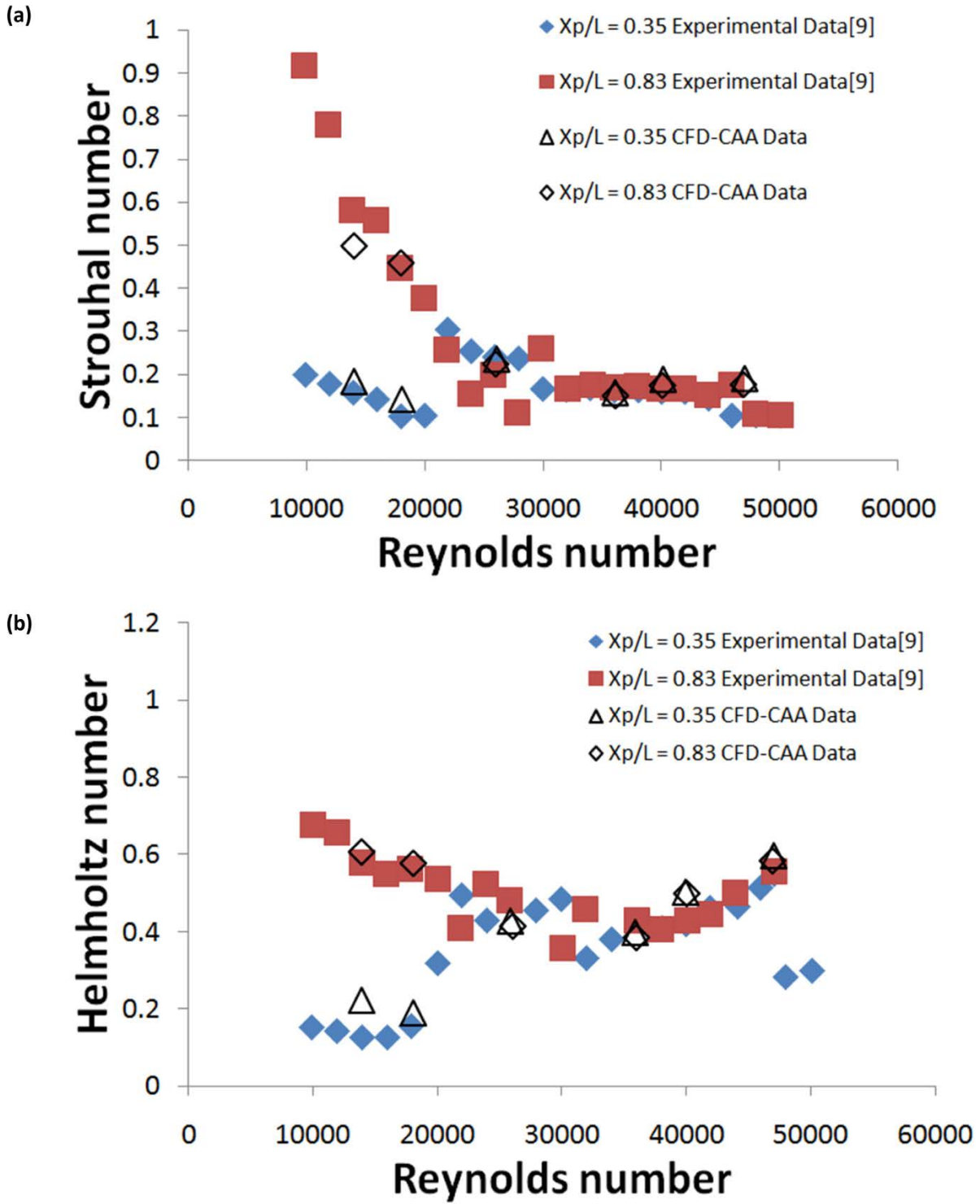

Figure 4. (a) Strouhal number (b)Helmholtz number at two points downstream of BFS for a range of Re

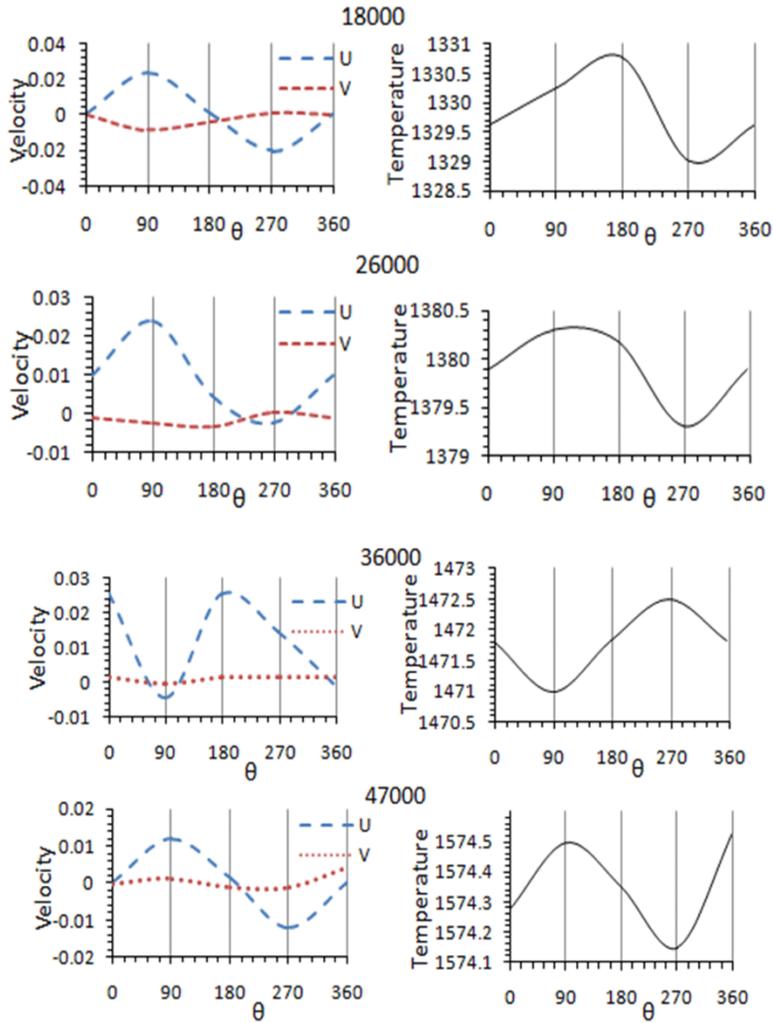

Figure 5. Reacting flow results for four different Re, lines are DES-EDM predictions: Phase average plots at primary vortex center U-Velocity (m/s - - - ), V-Velocity (m/s …..) and temperature (K —) for a cycle of oscillation.

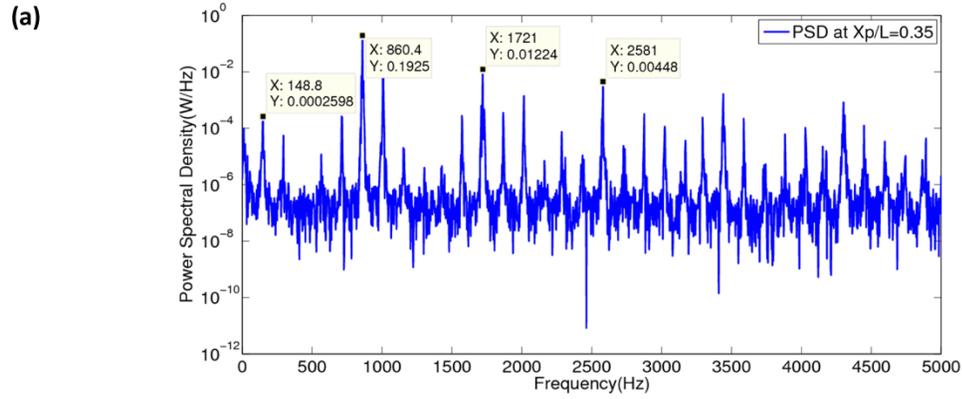

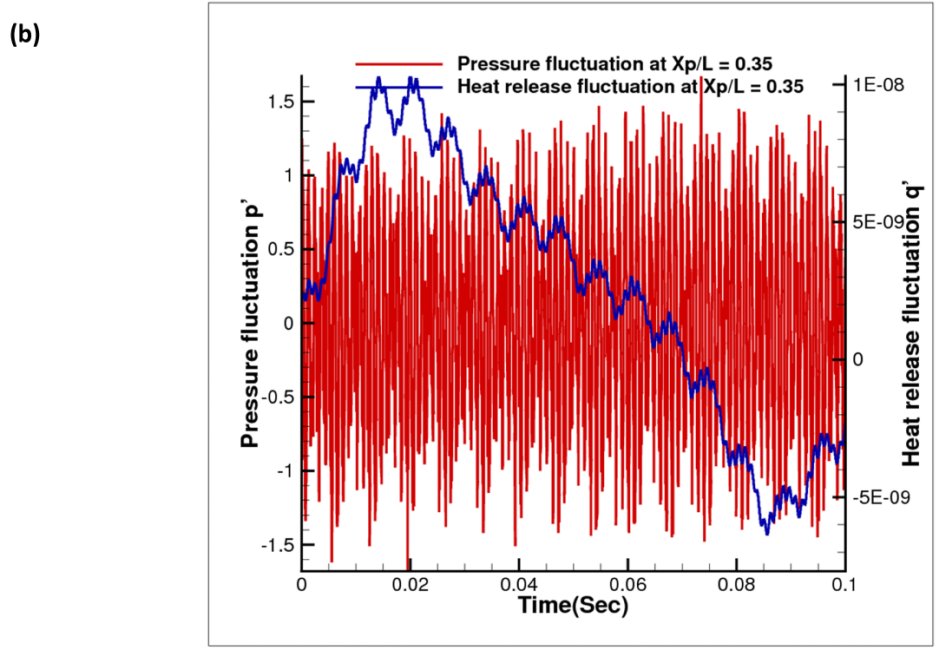

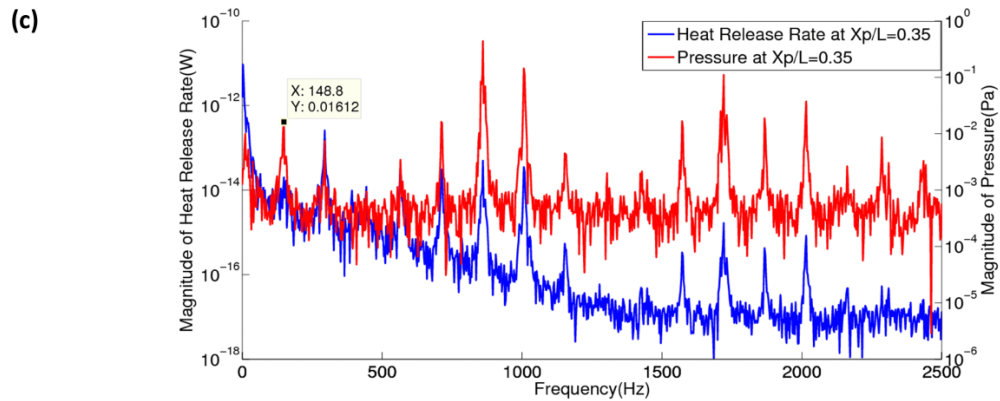

Figure 6. Reacting CAA results for Re= 47000 at Xp/L=0.35: (a) PSD vs Frequency (b) Pressure and hrr fluctuations w.r.t time (c) PSD of Pressure and hrr vs frequency

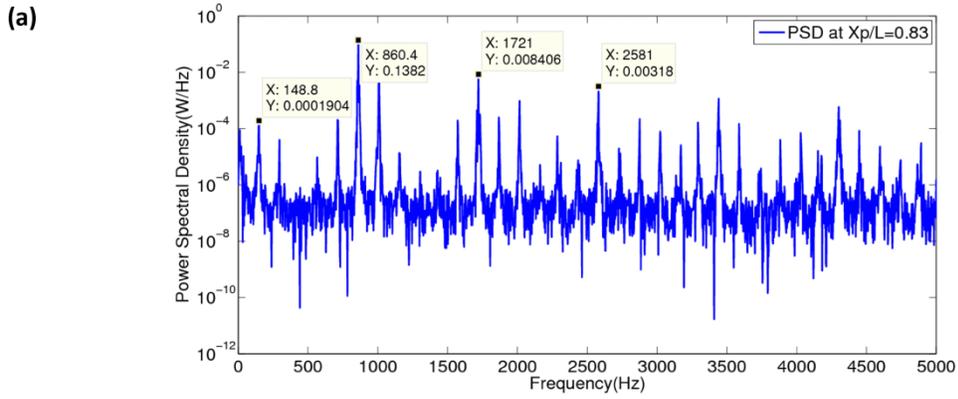

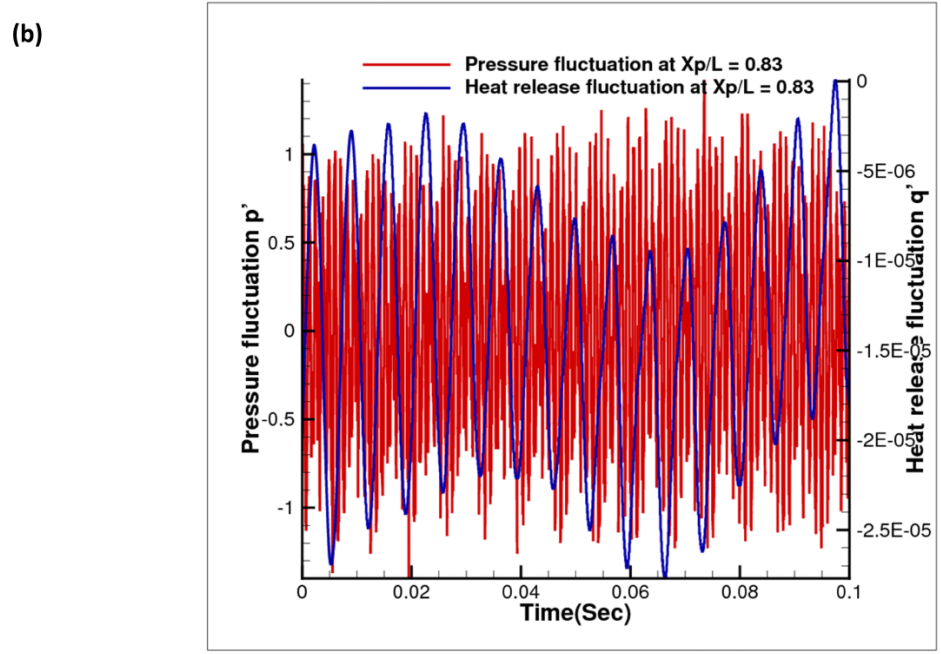

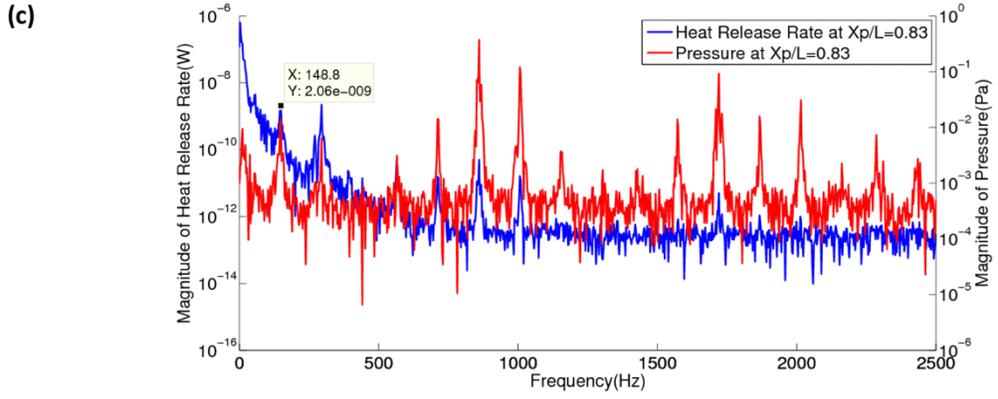

Figure 7. Reacting CAA results for Re= 47000 at Xp/L=0.83: (a) PSD vs Frequency (b)Pressure and hrr fluctuations w.r.t time (c) PSD of Pressure and hrr vs frequency

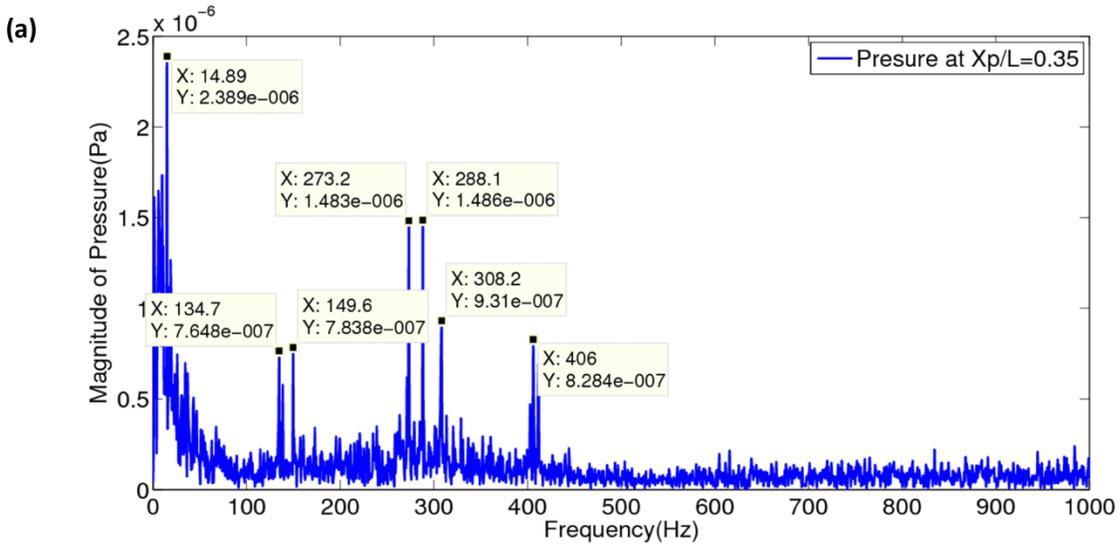

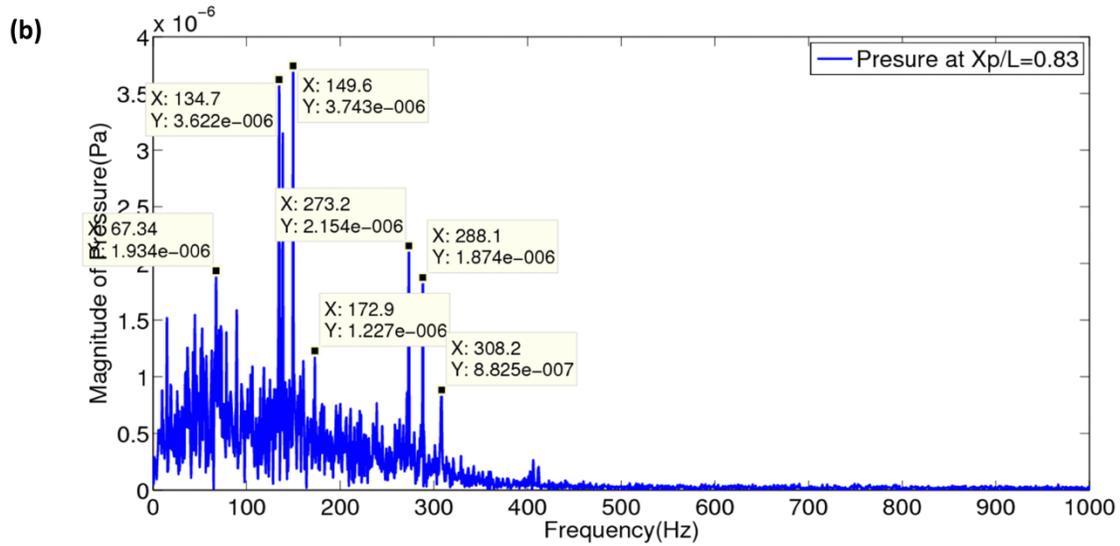

Figure 8. Non reacting CAA results for Re=47000. Magnitude of acoustic pressure vs Frequency at (a) Xp/L=0.35 and (b) Xp/L=0.83

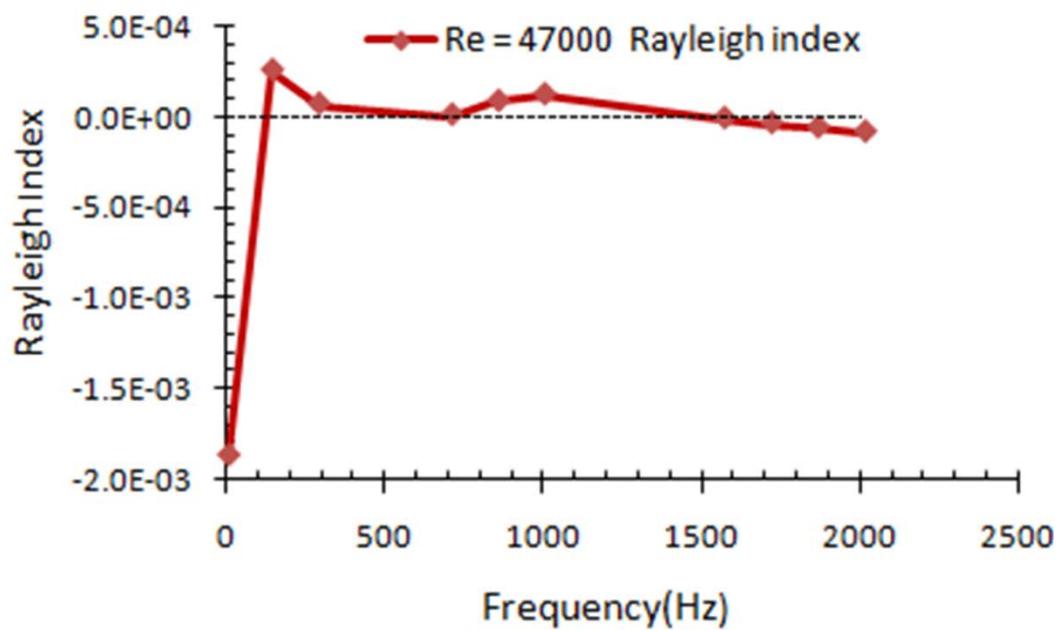

Figure 9.Rayleigh Index plot from reacting CAA data for Re=47000 at Xp/L=0.83

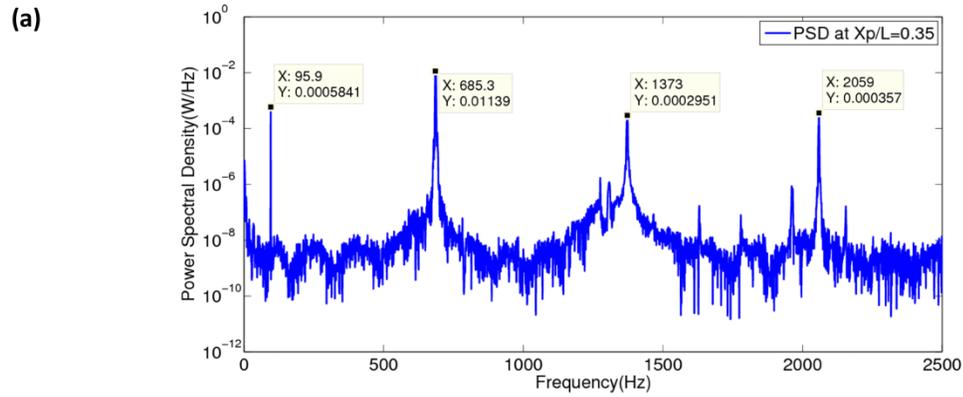

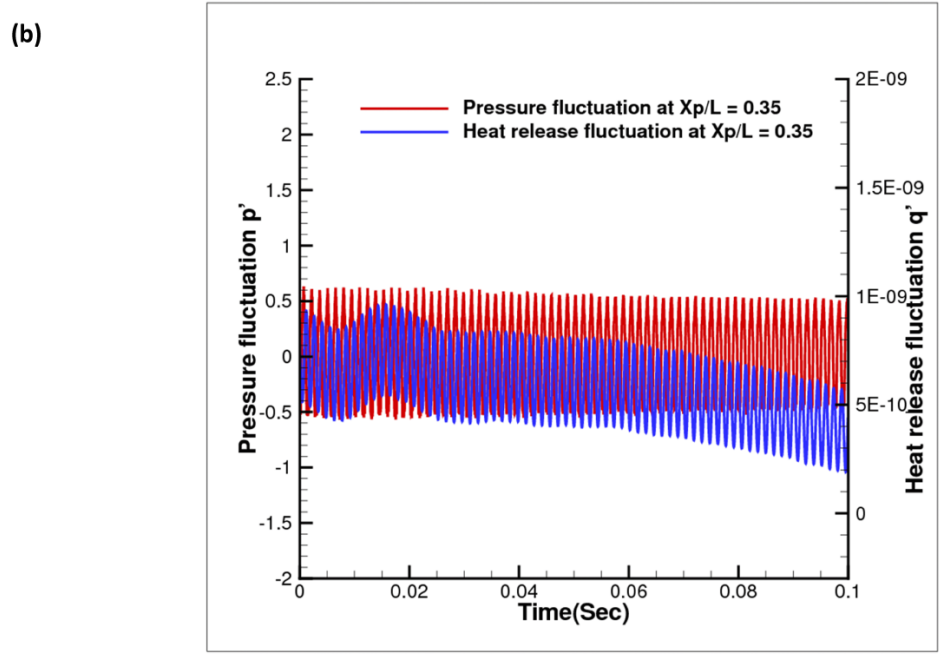

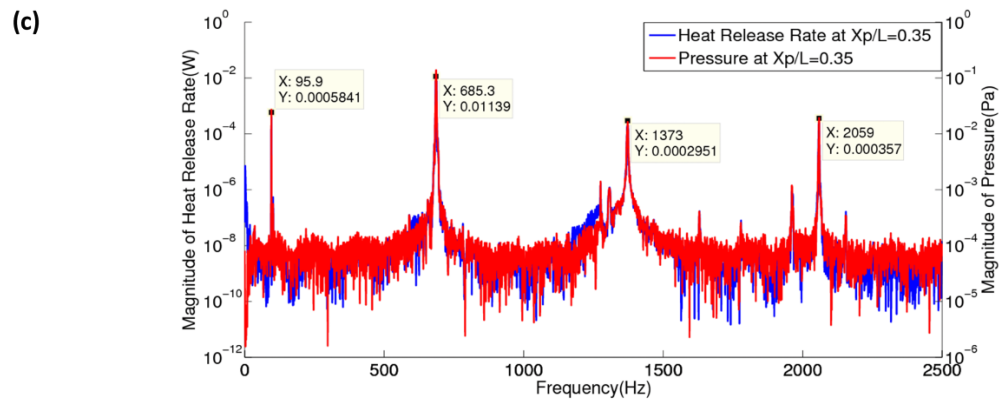

Figure 10. Reacting CAA results for Re=36000 at Xp/L=0.35: (a) PSD vs Frequency (b) Pressure and hrr fluctuations w.r.t time (c) PSD of Pressure and hrr vs frequency

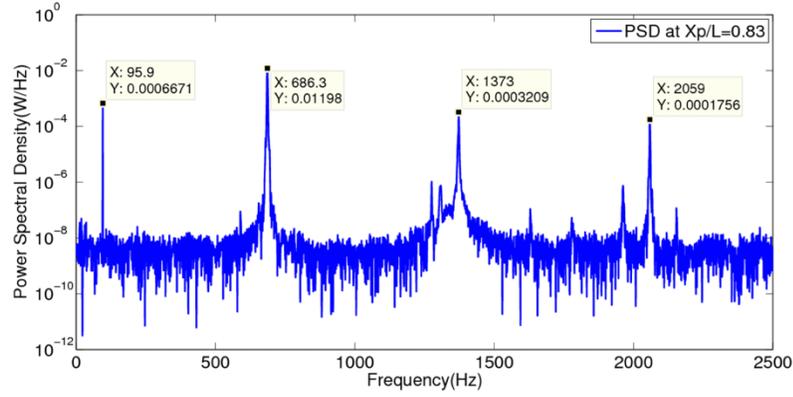

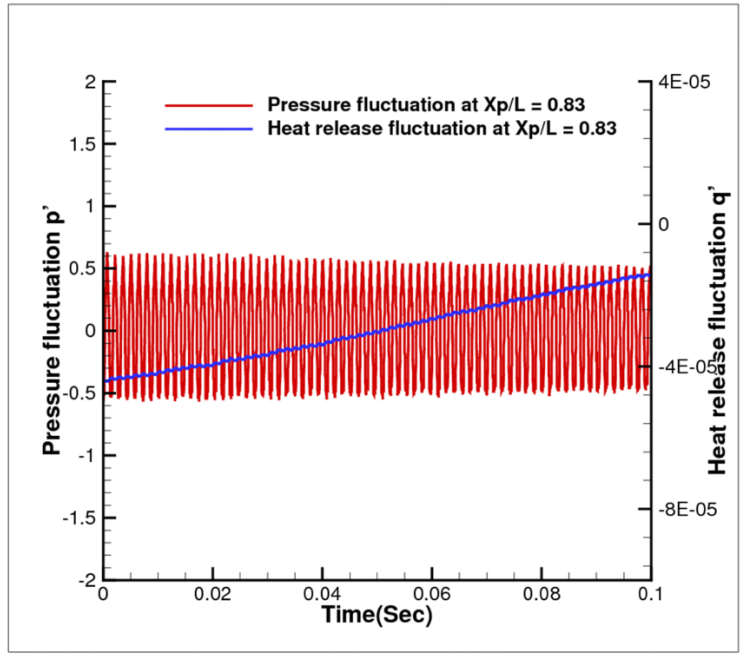

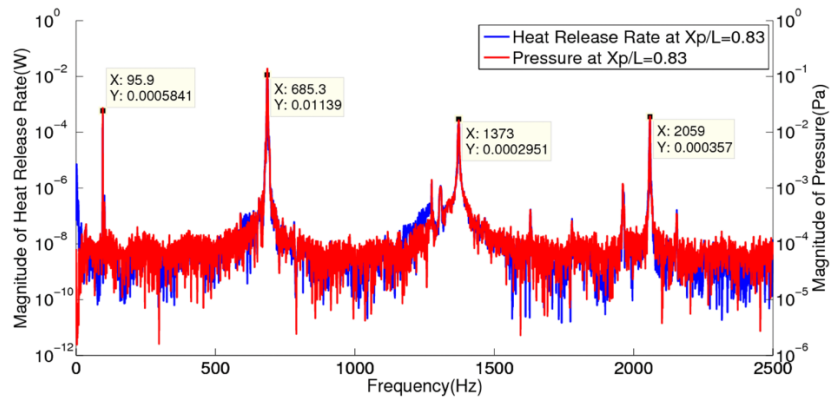

Figure 11. Reacting CAA results for Re=36000 at Xp/L=0.83: (a) PSD vs Frequency, (b) Pressure and hrr fluctuations w.r.t time (c) PSD of Pressure and hrr vs frequency

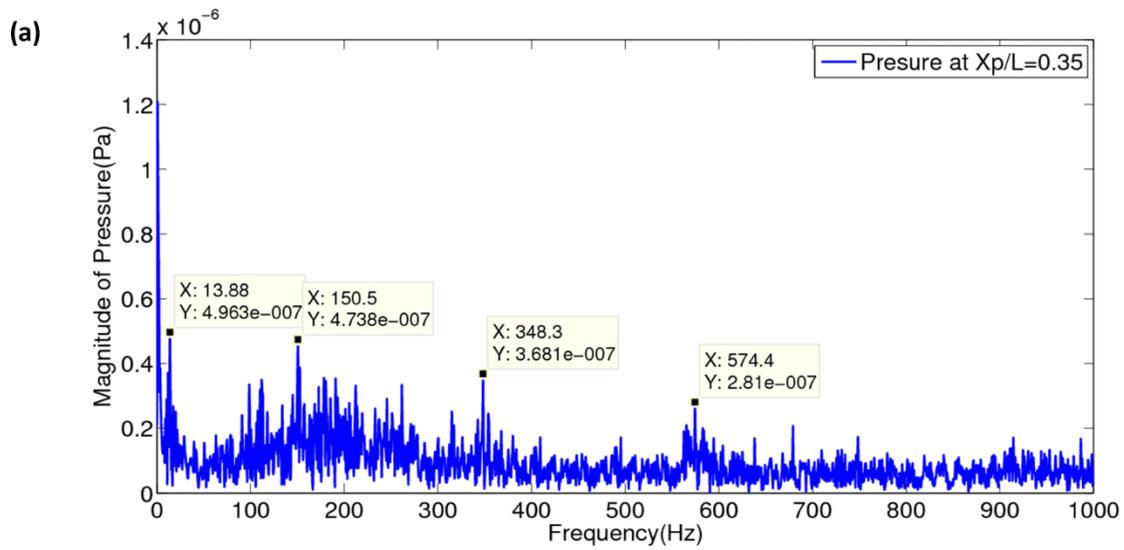

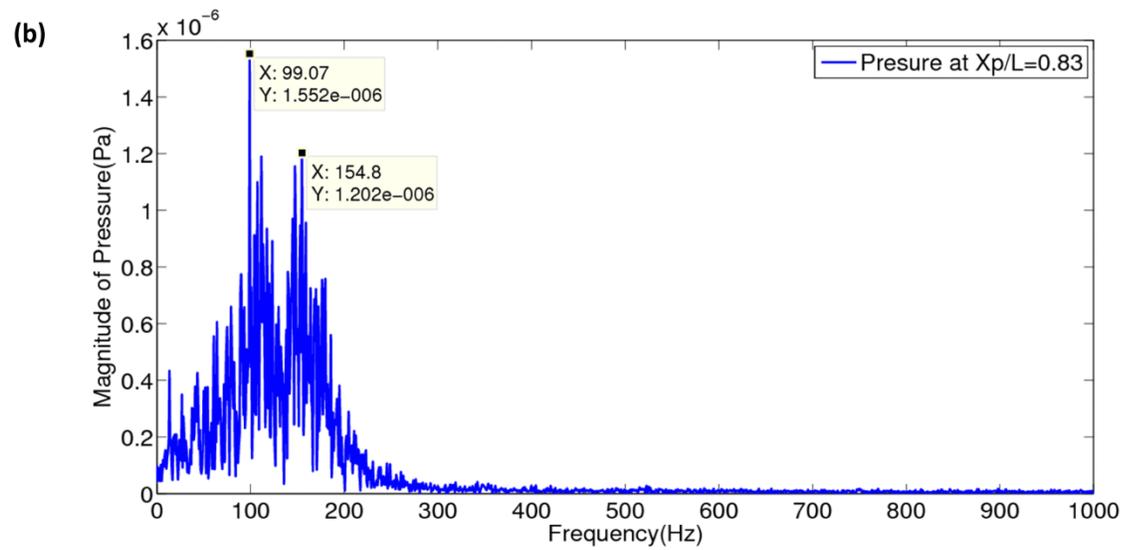

Figure 12. Non reacting CAA results for Re= 36000. Magnitude of acoustic pressure vs Frequency at (a) Xp/L=0.35 and (b) Xp/L=0.83

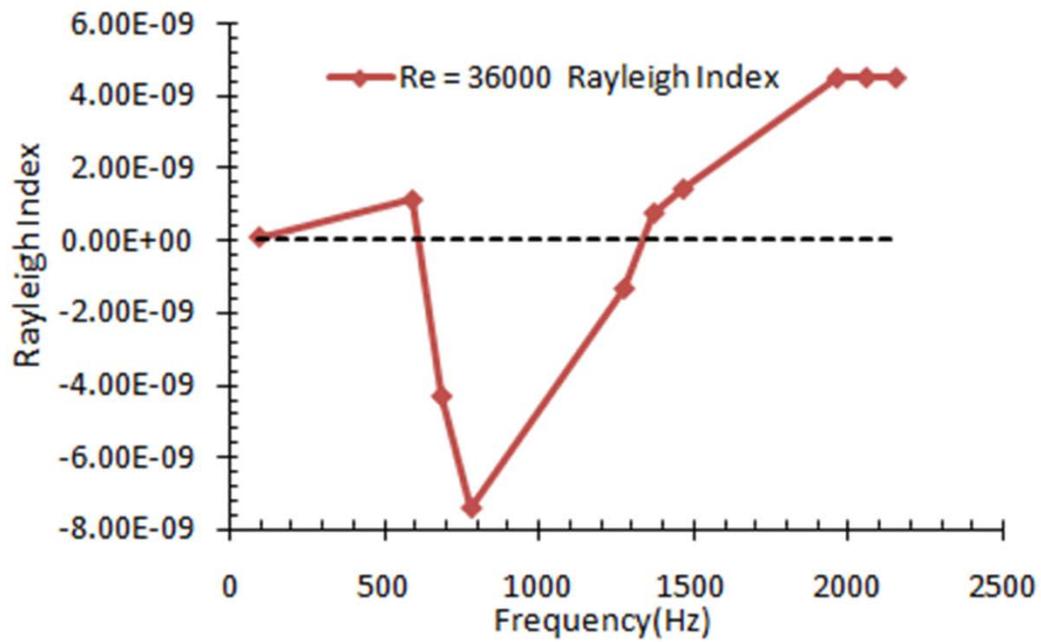

Figure 13.Rayleigh Index plot from reacting CAA data for Re=36000 at Xp/L=0.83

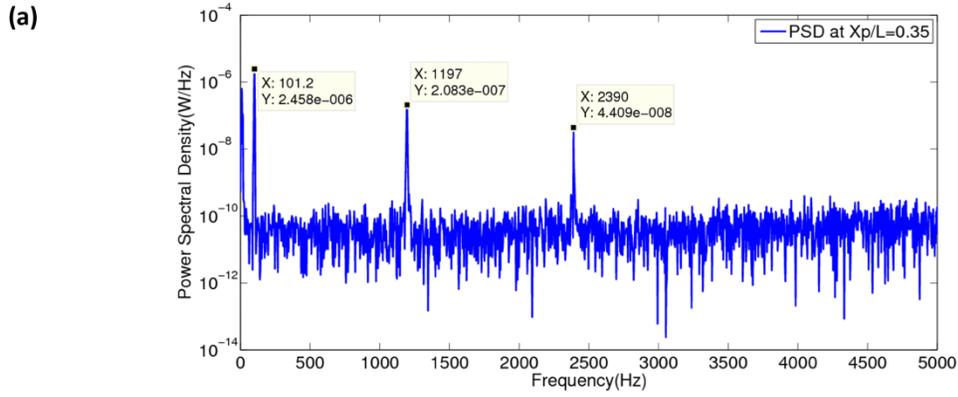

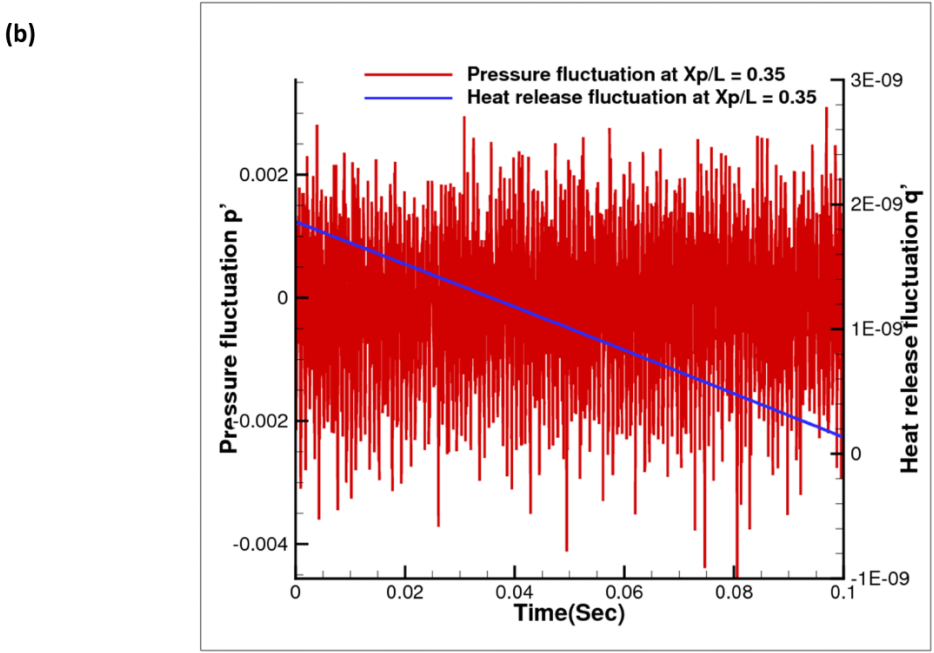

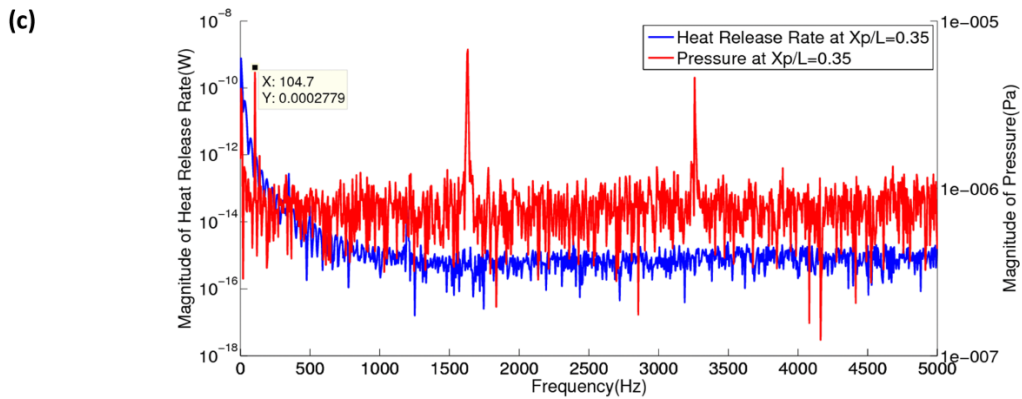

Figure 14. Reacting CAA results for Re=26000 at Xp/L=0.35: (a) PSD vs Frequency (b)Pressure and hrr fluctuations w.r.t time (c) PSD of Pressure and hrr vs frequency

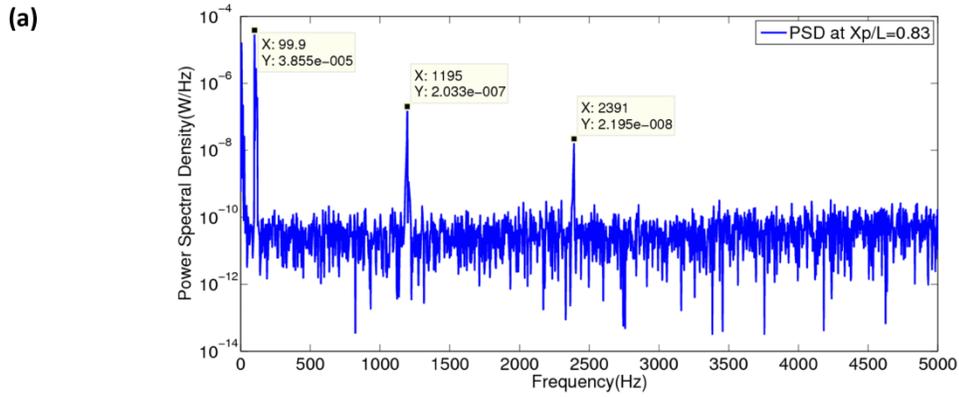

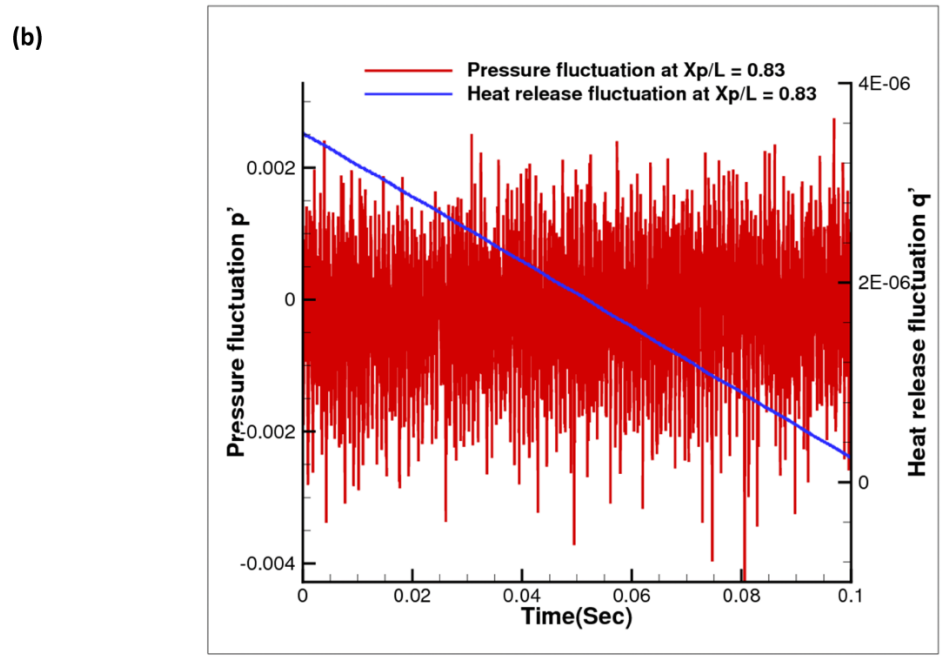

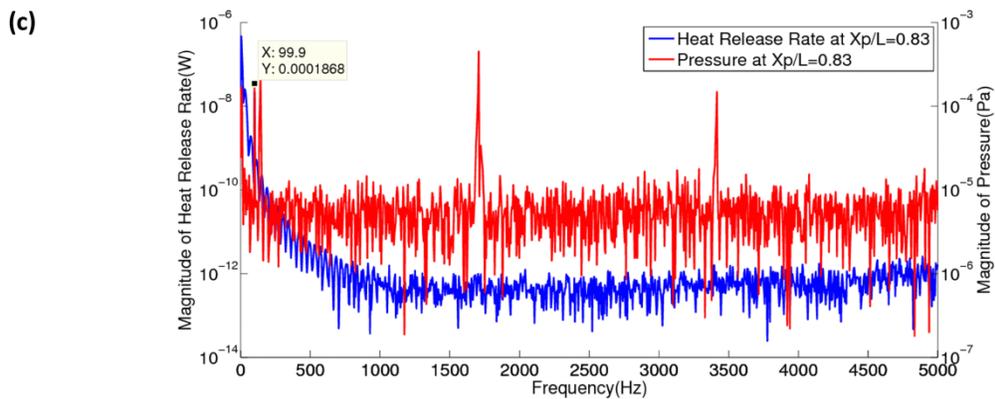

Figure 15. Reacting CAA results for Re=26000 at Xp/L=0.83: (a) PSD vs Frequency (b) Pressure and hrr fluctuations w.r.t time (c) PSD of Pressure and hrr vs frequency

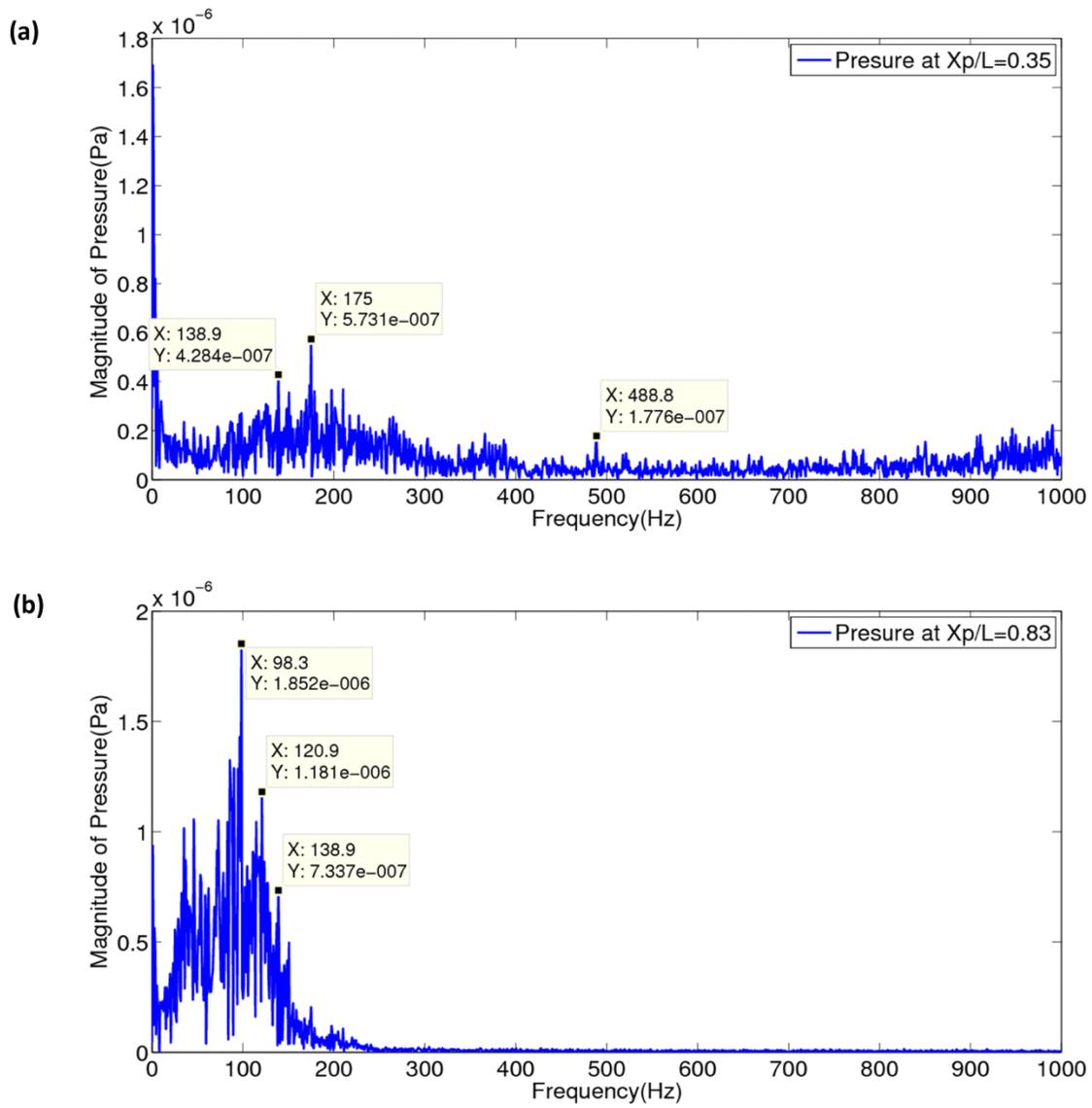

Figure 16. Non reacting CAA results for Re= 26000. Magnitude of acoustic pressure w.r.t Frequency at (a) Xp/L=0.35 and (b) Xp/L=0.83

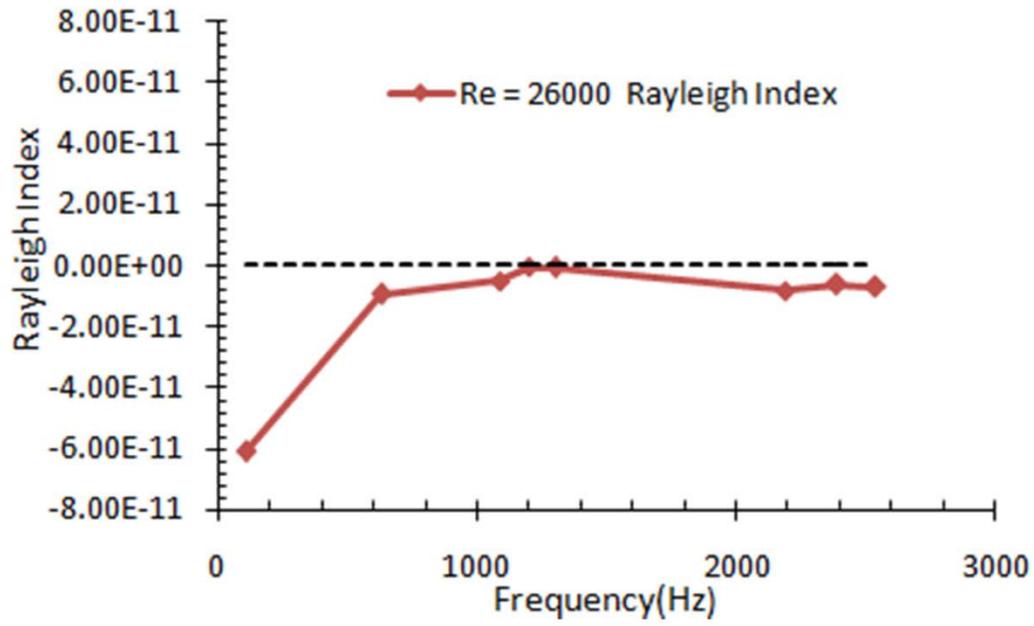

Figure 17.Rayleigh Index plot from reacting CAA data for Re=26000 at Xp/L=0.83

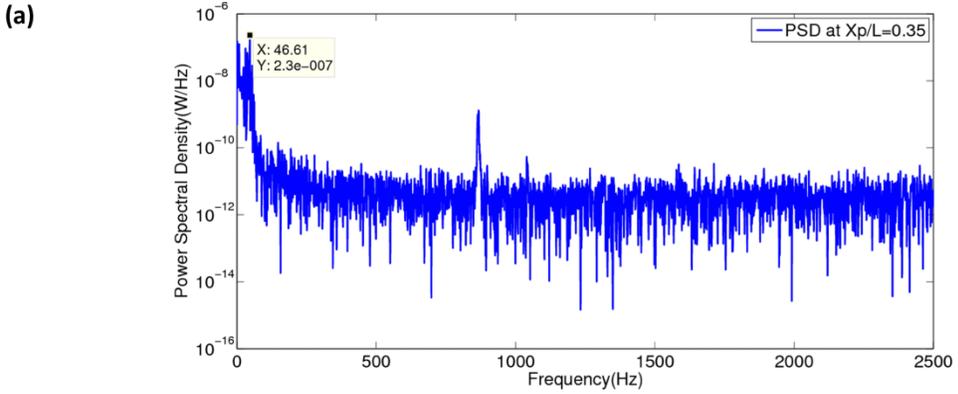

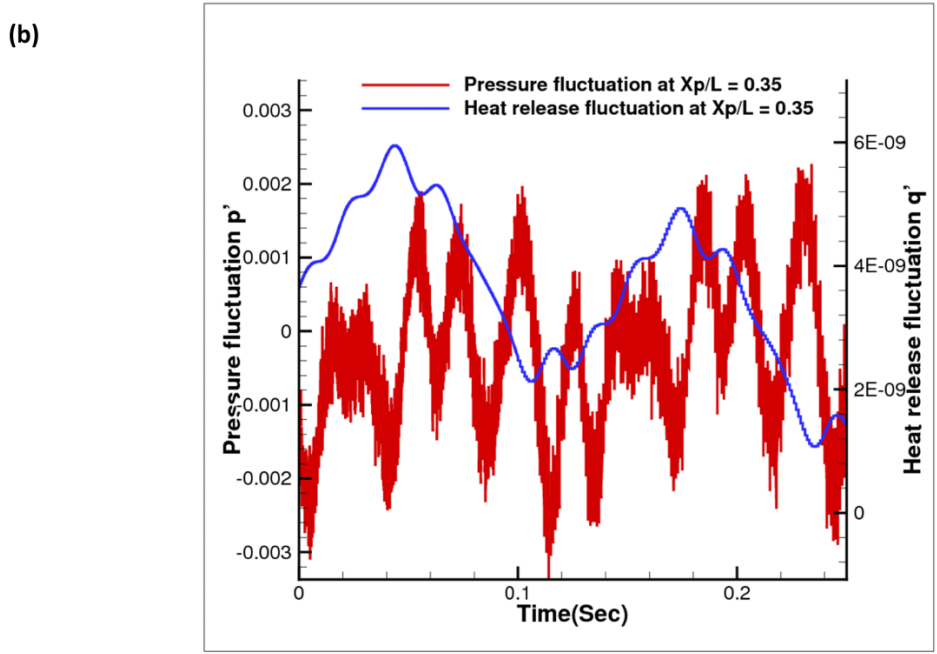

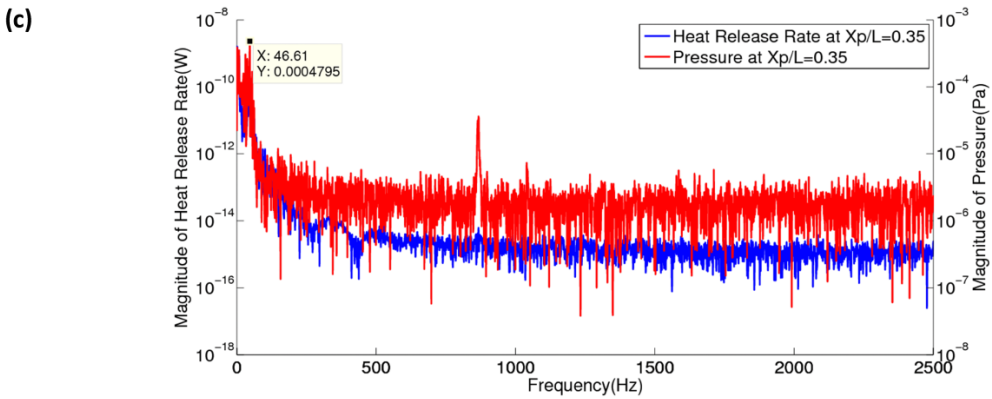

Figure 18. Reacting CAA results for Re= 18000 at Xp/L=0.35: (a) PSD vs Frequency (b)Pressure and hrr fluctuations w.r.t time (c) PSD of Pressure and hrr vs frequency

(a)

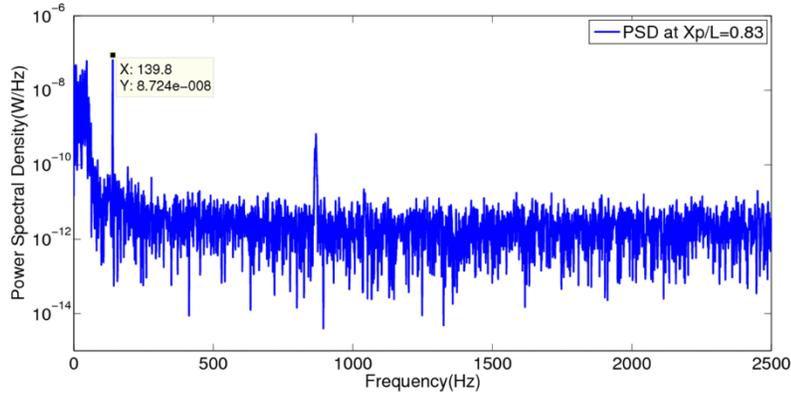

(b)

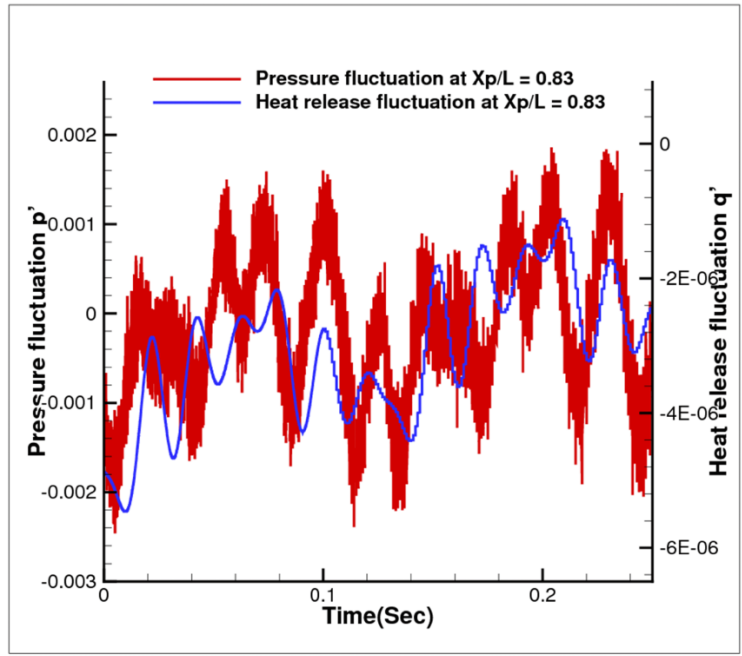

(c)

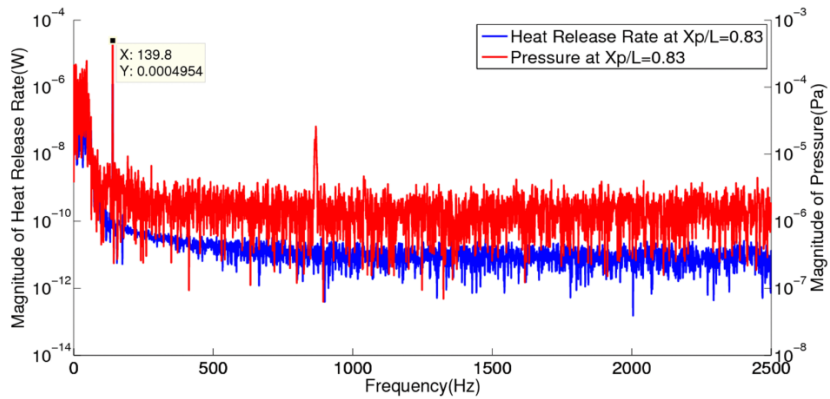

Figure 19. Reacting CAA results for Re=18000 at Xp/L=0.83 (a) PSD vs Frequency (b) Pressure and hrr fluctuations w.r.t time (c) PSD of Pressure and hrr vs frequency

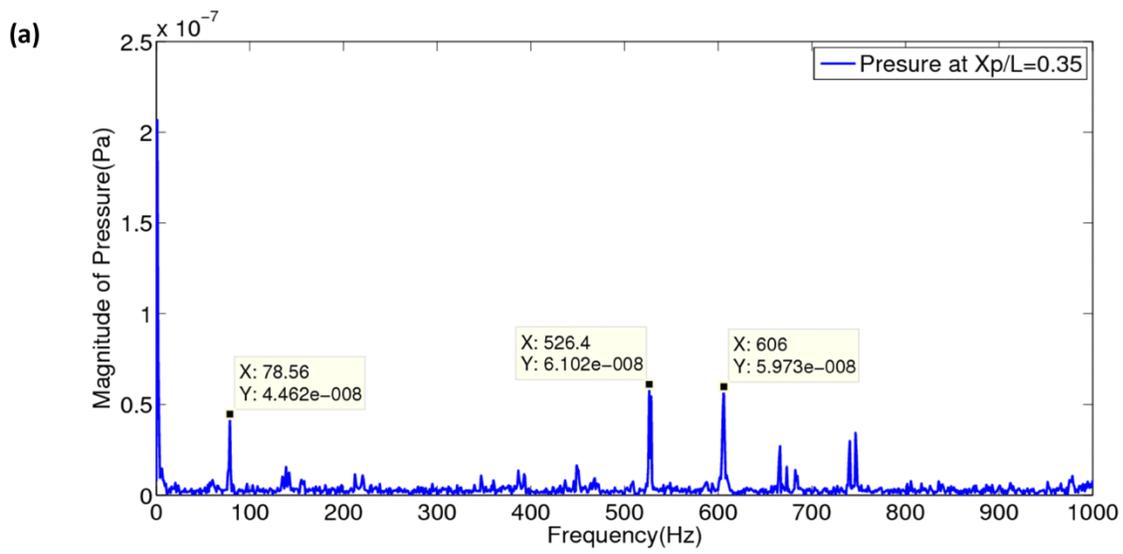

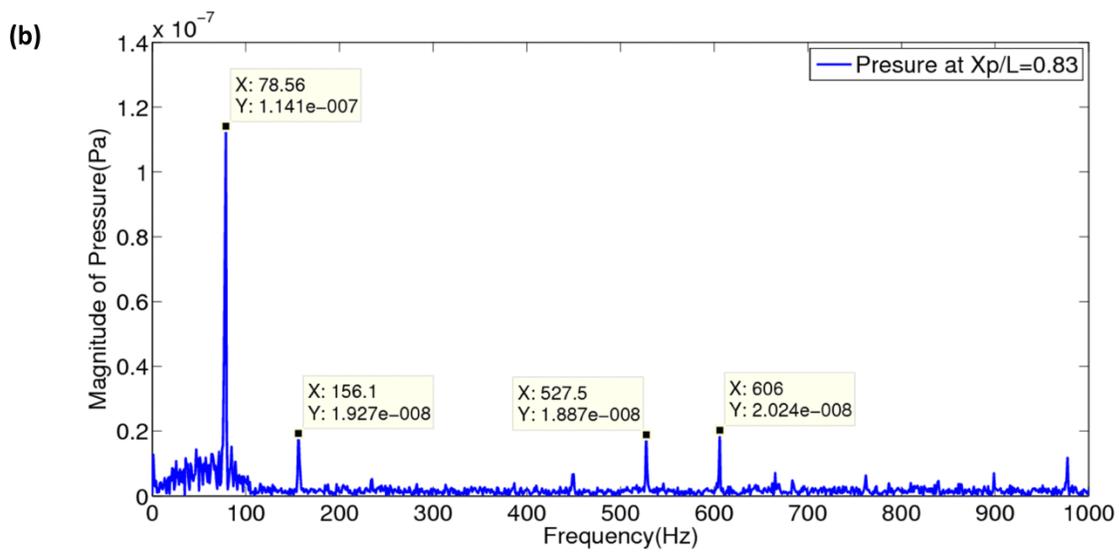

Figure 20. Non reacting CAA results for Re=18000. Magnitude of acoustic pressure w.r.t Frequency at (a) Xp/L=0.35 and (b) Xp/L=0.83

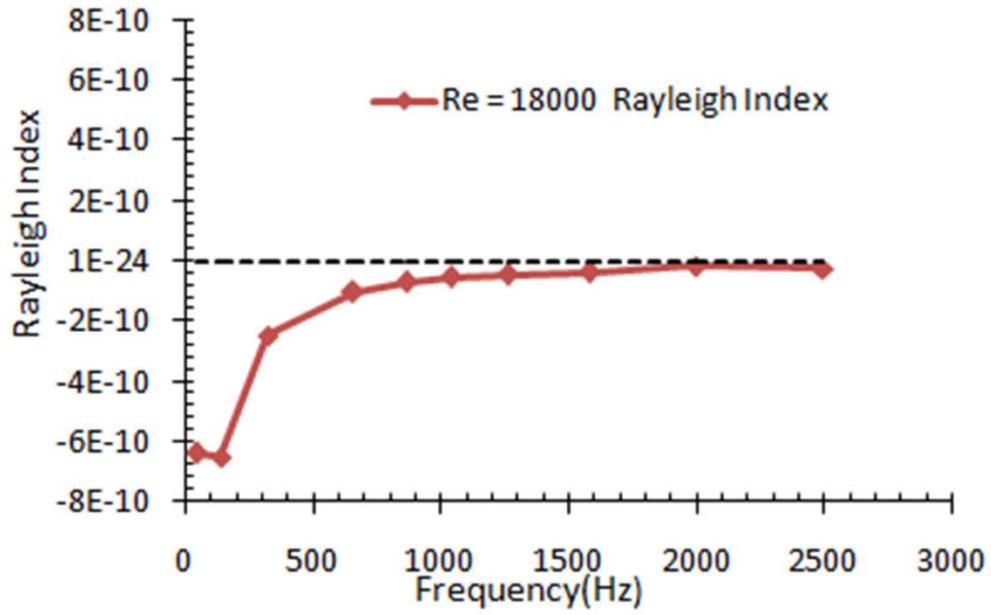

Figure 21.Rayleigh Index plot from reacting CAA data for R=18000 at Xp/L=0.83